\documentclass[letterpaper,twocolumn,10pt]{article}
\usepackage{usenix2019_v3}

\usepackage[utf8]{inputenc}

\usepackage[kerning,spacing]{microtype} 
\usepackage{flushend} 

\usepackage{cite}               
\usepackage{breakurl}           
\usepackage{xcolor}             
\PassOptionsToPackage{hyphens}{url}\usepackage[hyperfootnotes=false]{hyperref}         
\hypersetup{                    
  colorlinks,
  linkcolor={green!80!black},
  citecolor={red!70!black},
  urlcolor={blue!70!black}
}

\usepackage{paralist}
\usepackage{algorithm}
\usepackage[noend]{distribalgo}
\usepackage{graphicx}
\usepackage{booktabs}
\usepackage{refcount}
\usepackage{totpages}
\usepackage{xspace}
\usepackage[font=small,format=plain,labelfont=bf,textfont=bf]{caption}
\usepackage{float}
\usepackage{amsmath}
\usepackage{amssymb}
\usepackage{amsthm}
\usepackage{pifont}
\usepackage{colortbl}

\newif\ifsubmission
\submissiontrue

\newcommand{\cmark}{}
\newcommand{\xmark}{\cellcolor{gray!25}}
\newcommand{\smark}{\ding{218}}
\newcommand{\xxmark}{\cellcolor{red!25}}

\newcommand{\mv}[1]{\emph{#1}}

\newcommand{\R}{\mathbb{R}}

\newcommand{\system}{SwitchML\xspace}

\newcommand{\smartparagraph}[1]{\smallskip\noindent {\bf #1}}

\ifsubmission
\newcommand{\asnote}[1]{}
\newcommand{\mcnote}[1]{}
\newcommand{\prnote}[1]{}
\newcommand{\aknote}[1]{}
\newcommand{\cknote}[1]{}
\newcommand{\panosnote}[1]{}
\newcommand{\dpnote}[1]{}
\else
\newcommand{\asnote}[1]{\textit{\textcolor{orange}{[amedeo]: #1}}} 
\newcommand{\mcnote}[1]{\textit{\textcolor{blue}{[marco]: #1}}} 
\newcommand{\prnote}[1]{\textit{\textcolor{red}{[peter]: #1}}} 
\newcommand{\aknote}[1]{\textit{\textcolor{magenta}{[arvind]: #1}}} 
\newcommand{\cknote}[1]{\textit{\textcolor{brown}{[chang]: #1}}} 
\newcommand{\panosnote}[1]{\textit{\textcolor{green}{[panos]: #1}}} 
\newcommand{\dpnote}[1]{\textit{\textcolor{cyan}{[dan]: #1}}} 
\fi

\usepackage[page]{totalcount}

\begin{document}

\date{}

\title{\Large \bf Scaling Distributed Machine Learning with In-Network Aggregation}

 \author{
 {\rm Amedeo Sapio}\thanks{Equal contribution.}\\
 KAUST
 \and
 {\rm Marco Canini}\footnotemark[1]\\
 KAUST
 \and
 {\rm Chen-Yu Ho}\\
 KAUST
 \and
 {\rm Jacob Nelson}\\
 Microsoft
\end{tabular}\\
\begin{tabular}[t]{c}
 {\rm Panos Kalnis}\\
 KAUST
 \and
 {\rm Changhoon Kim}\\
 Barefoot Networks
 \and
 {\rm Arvind Krishnamurthy}\\
 University of Washington
\end{tabular}\\
\begin{tabular}[t]{c}
 {\rm Masoud Moshref}\\
 Barefoot Networks
 \and
 {\rm Dan R. K. Ports}\\
 Microsoft
 \and
 {\rm Peter Richt\'arik}\\
 KAUST
 } 

\maketitle

\begin{abstract}

Training machine learning models in parallel is an increasingly important workload. We accelerate distributed parallel training by designing a communication primitive that uses a programmable switch dataplane to execute a key step of the training process. Our approach, \system, reduces the volume of exchanged data by aggregating the model updates from multiple workers in the network. We co-design the switch processing with the end-host protocols and ML frameworks to provide an efficient solution that speeds up training by up to 5.5$\times$ for a number of real-world benchmark models.

\end{abstract}

\thispagestyle{plain}

\section{Introduction}

Today's machine learning solutions' remarkable success derives from the ability to build increasingly sophisticated models on increasingly large data sets. To cope with the resulting increase in training time, ML practitioners use distributed training~\cite{distbelief, tensorflow}. Large-scale clusters use hundreds of nodes, each equipped with multiple GPUs or other hardware accelerators (e.g., TPUs~\cite{tpu}), to run training jobs on tens of workers that take many hours or days.

Distributed training is increasingly a \emph{network-bound} workload. To be clear, it remains computationally intensive. But the last seven years have brought a $62\times$ improvement in compute performance~\cite{phub2018,nvidia-perf}, thanks to GPUs~\cite{nvidia-a100} and other hardware accelerators~\cite{tpu,graphcore,cerebras,habana}). Cloud network deployments have found this pace hard to match, skewing the ratio of computation to communication towards the latter. Since parallelization techniques like mini-batch stochastic gradient descent (SGD) training~\cite{applied-ML-FB, gpu_sched} alternate computation with synchronous model updates among workers, network performance now has a substantial impact on training time.

Can a new type of accelerator \emph{in the network} alleviate the network bottleneck? We demonstrate that an \emph{in-network aggregation} primitive can accelerate distributed ML workloads, and can be implemented using programmable switch hardware~\cite{rmt,tofino}. Aggregation reduces the amount of data transmitted during synchronization phases, which increases throughput, diminishes latency, and speeds up training time.

Building an in-network aggregation primitive using programmable switches presents many challenges.  First, the per-packet processing capabilities are limited, and so is on-chip memory. 
We must limit our resource usage so that the switch can perform its primary function of conveying packets.  Second, the computing units inside a programmable switch operate on integer values, whereas ML frameworks and models operate on floating-point values.  Finally, the in-network aggregation primitive is an all-to-all primitive, unlike traditional unicast or multicast communication patterns. As a result, in-network aggregation requires mechanisms for synchronizing workers and detecting and recovering from packet loss.

We address these challenges in \system, showing that it is indeed possible for a programmable network device to perform in-network aggregation at line rate. \system is a co-design of in-switch processing with an end-host transport layer and ML frameworks. It leverages the following insights. First, aggregation involves a simple arithmetic operation, making it amenable to parallelization and pipelined execution on programmable network devices.  We decompose the parameter updates into appropriately-sized chunks that can be individually processed by the switch pipeline. Second, aggregation for SGD can be applied separately on different portions of the input data, disregarding order, without affecting the correctness of the final result. We tolerate packet loss through the use of a light-weight switch scoreboard mechanism and a retransmission mechanism driven solely by end hosts, which together ensure that workers operate in lock-step without any decrease in switch aggregation throughput. Third, ML training is robust to modest approximations in its compute operations.  
We address the lack of floating-point support in switch dataplanes by having the workers scale and convert floating-point values to fixed-point
using an adaptive scaling factor with negligible approximation loss.

\system integrates with distributed ML frameworks, such as PyTorch and TensorFlow, to accelerate their communication, and enable efficient training of deep neural networks (DNNs). Our initial prototype targets a \emph{rack-scale architecture}, where a single switch centrally aggregates parameter updates from serviced workers. Though the single switch limits scalability, we note that commercially-available programmable switches can service up to 64 nodes at 100 Gbps or 256 at 25 Gbps.  As each worker is typically equipped with multiple GPUs, this scale is sufficiently large to push the statistical limits of SGD \cite{gpu_sched, imagenet-1hour, keskar2016large, youspeeding}.

We show that \system's in-network aggregation yields end-to-end improvements in training performance of up to $5.5\times$ for popular DNN models.
Focusing on a communication microbenchmark,
compared to the best-in-class collective library NCCL~\cite{nvidia-nccl},
\system is up to 2.9$\times$ faster than NCCL with RDMA and 9.1$\times$ than NCCL with TCP.
While the magnitude of the performance improvements is dependent on the neural network architecture and the underlying physical network speed, it is greater for models with smaller compute-to-communication ratios -- good news for future, faster DNN training accelerators.

Our approach is not tied to any particular ML framework; we have integrated \system with Horovod~\cite{horovod}, which supports several popular toolkits like TensorFlow and PyTorch. We plan to release \system as open-source software.

\vspace{-0.3em}

\section{Network bottlenecks in ML training}

In the distributed setting, ML training yields a high-performance networking problem, which we highlight below after reviewing the traditional ML training process.

\subsection{Training and all to all communication}
\label{sec:comm}

Supervised  ML problems, including logistic regression, support vector machines and  deep learning, are typically solved by iterative algorithms such as stochastic gradient descent (SGD)~\cite{RobbinsMonro:1951, NemYudin1983book, Nemirovski-Juditsky-Lan-Shapiro-2009} or one of its many variants (e.g., using momentum, mini-batching, importance sampling, preconditioning, variance reduction)~\cite{shai_book}. A common approach to scaling to large models and datasets is data-parallelism, where the input data is partitioned across workers.\footnote{In this paper, we do not consider model-parallel training \cite{Hydra, Hydra2}, although that approach also requires efficient networking. Further, we focus exclusively on widely-used distributed synchronous SGD~\cite{applied-ML-FB, tensorflow}.} 
Training in a data-parallel, synchronized fashion on $n$ workers can be seen as learning a model $x\in \R^d$ over input/training data $D$ by performing iterations of the form
$x^{t+1} = x^{t} + \sum_{i=1}^{n} \Delta(x^{t}, D_i^t),$
where $x^t$ is a vector of model parameters\footnote{In applications, $x$ is typically a 1, 2, or 3 dimensional tensor. To simplify notation, we assume its entries are vectorized into one $d$ dimensional vector.} at iteration $t$, $\Delta(\cdot, \cdot)$ is the model update function\footnote{We abstract learning rate (step size) and model averaging inside $\Delta$.} and $D_i^t$ is the data subset used at worker $i$ during that iteration. 

The key to data parallelism is that each worker $i$, in parallel, locally computes the
update $\Delta(x^{t}, D_i^t)$ to the model parameters based on the
current model $x^t$ and a mini-batch, i.e., a subset of the local data $D_i^t$.
Typically, a model update contributed by worker $i$ is a multiple of the stochastic gradient of the loss function with respect to the current model parameters $x^t$ computed across a mini-batch of training data, $D_i^t$.
Subsequently, workers communicate their updates, which are aggregated
($\sum$) and added to $x^{t}$ to form the model parameters of the next
iteration.  Importantly, each iteration acts only on a
mini-batch of the training data. It requires many iterations
to progress through the entire dataset, which constitutes a training
epoch. A typical training job requires multiple epochs, reprocessing
the full training data set, until the model achieves acceptable error
on a validation set.

From a networking perspective, the challenge is that data-parallel SGD
requires computing the sum of model updates across all workers after
every iteration. Each model update has as many parameters as the model
itself, so they are often in 100s-of-MB or GB range. And
their size is growing exponentially: today's largest models exceed
32~GB~\cite{turing-nlg}. These aggregations need to be performed
frequently, as increasing the mini-batch size hurts
convergence~\cite{revisiting-small-batch}.
Today's ML toolkits implement this communication phase in one of two ways:

\smartparagraph{The parameter server (PS) approach.} In this approach,
workers compute model updates and send them to \emph{parameter
  servers}. These servers, usually dedicated machines, aggregate
updates to compute and distribute the new model parameters. To prevent
the PS from becoming a bottleneck, the model is sharded over multiple
PS nodes.

\smartparagraph{The all-reduce approach.}
An alternate 
approach uses the workers to run an all-reduce algorithm -- a collective communication technique common in high-performance computing --
to combine model updates. The workers communicate over an overlay
network. A \emph{ring} topology~\cite{intercom}, where each worker
communicates to the next neighboring worker on the ring, is common
because it is bandwidth-optimal (though its latency grows with the number of workers)~\cite{Patarasuk2009117}. 
\emph{Halving and doubling} uses a binary tree
topology~\cite{opt-mpi} instead.

\subsection{The network bottleneck}
\label{sec:network-bound}
Fundamentally, training alternates compute-intensive phases with
communication-intensive model update synchronization. Workers produce
intense bursts of traffic to communicate their model updates, whether
it is done through a parameter server or all-reduce, and training
stalls until it is complete.

Recent studies have shown that performance bottleneck in distributed
training is increasingly shifting from compute to
communication~\cite{phub2018}. This shift comes from two sources. The
first is a result of advances in GPUs and
other compute accelerators. 
For example, the recently released NVIDIA
A100 offers $10\times$ and $20\times$ performance improvements for
floating-point and mixed-precision calculations,
respectively~\cite{nvidia-a100} compared to its predecessor, the V100
-- released just 2.5 years previously. This pace far exceeds advances
in network bandwidth: a $10\times$ improvement in Ethernet speeds
(from 10 Gbps to 100 Gbps) required 8 years to standardize. 

Second, the ratio of communication to computation in the workload
itself has shifted. The current trend towards ever-larger DNNs
generally exacerbates this issue. However, this effect is highly
application-dependent. In popular ML toolkits, communication and
computation phases can partially overlap. Since back-prop
proceeds incrementally, communication can start as soon as the
earliest partial results of back-prop are available. The
effectiveness of this technique depends on the structure of the DNN.
For DNNs with large initial layers, its effectiveness is marginal,
because there is little to no opportunity to overlap communication
with computation.

\smartparagraph{When is the network a bottleneck?} To answer this
quantitatively, we profile the training of 8 common DNNs on a cluster
with 8 workers using NVIDIA P100 GPUs.
To precisely factor out the contribution of communication to the processing time of a mini-batch, we emulate communication time at 10 Gbps or 100 Gbps Ethernet assuming transmission at line rate.
We record the network-level events,
which allows us to report the fraction of time spent in communication as well as
how much can overlap with computation (Table~\ref{tab:dnns}).
At 10 Gbps, all but three training jobs spend more than 50\% of their
time in communication, usually with little computation-phase
overlap. These workloads benefit greatly from 100 Gbps networking, but even so communication remains an average of 23\% of
batch processing time.

\smartparagraph{What happens when GPUs become faster?} Our profile
uses P100 GPUs, now two generations old. Faster GPUs would reduce the
computation time, increasing the relative communication fraction. Our
measurement of non-overlappable communication time allows us to
determine the scaling factor $\alpha$ applied to GPU computation time at which point the network is
saturated. There is still some speedup beyond an $\alpha \times$
faster GPU, but it is limited to the initial phase, before
communication begins. Note $\alpha < 4$ for half the workloads,
suggesting that network performance will be a serious issue when using
the latest GPUs with a 100~Gbps network.

\begin{table}[t]
\resizebox{\columnwidth}{!}{%
\setlength\tabcolsep{2.5pt} 
\begin{tabular}{lrrrrrrr}
\toprule
Model     & Size & \multicolumn{1}{l}{Batch} & \multicolumn{2}{c}{10 Gbps}                                           & \multicolumn{3}{c}{100 Gbps}                                                                  \\
\cmidrule(lr){4-5}\cmidrule(lr){6-8}
          &  [MB] & size      & \multicolumn{1}{l}{Batch {[}ms{]}} & \multicolumn{1}{l}{Comm {[}\%{]}} & \multicolumn{1}{l}{Batch {[}ms{]}} & \multicolumn{1}{l}{Comm {[}\%{]}} & $\alpha$ \\
\midrule
DeepLight & 2319     & $2^{13}$ & 2101 $\pm$ 1.4     & 97\% (2\%)      & 258 $\pm$ 0.4      & 79\% (20\%)     & 1.0                   \\
LSTM      & 1627     & 64                    & 1534 $\pm$ 8.3     & 94\% (10\%)     & 312 $\pm$ 6.8      & 46\% (56\%)     & 1.5                   \\
BERT      & 1274      & 4                     & 1677 $\pm$ 7.1     & 67\% (3\%)      & 668 $\pm$ 3.1      & 17\% (35\%)     & 3.5                   \\
VGG19     & 548      & 64                    & 661 $\pm$ 1.9      & 73\% (67\%)     & 499 $\pm$ 1.1      & 10\% (99\%)     & 6.7                   \\
UGATIT    & 511      & 2                     & 1612 $\pm$ 2.5     & 28\% (84\%)     & 1212 $\pm$ 3.5     & 4\% (99\%)      & 17.6                  \\
NCF       & 121     & $2^{17}$ & 149 $\pm$ 0.6      & 72\% (4\%)      & 46 $\pm$ 0.1       & 23\% (27\%)     & 1.2                   \\
SSD       & 98       & 16                    & 293 $\pm$ 0.6      & 26\% (99\%)     & 293 $\pm$ 1.6      & 3\% (99\%)      & 15.2                  \\
ResNet-50 & 87       & 64                    & 299 $\pm$ 10.9     & 29\% (67\%)     & 270 $\pm$ 1.2      & 3\% (94\%)      & 19.8                 \\
\bottomrule
\end{tabular}
}
  \caption{Profile of benchmark DNNs. ``Batch [ms]'' reports the
    average batch processing time and its standard deviation. ``Comm''
    reports the proportion of communication activity as \% of batch
    time. The figure in parentheses is the percentage of \emph{that}
    time that overlaps with computation.
For example, DeepLight at 10 Gbps spends 97\% of its batch time in communication; only 2\% of this 97\% communication overlaps with computation. The table lists a scaling factor for an hypothetical $\alpha\times$ faster GPU that implies communication is contiguous and saturates the 100~Gbps bandwidth once communication begins.}
  \label{tab:dnns}
\end{table}

\section{In-network aggregation}
\label{sec:ina}

We propose an alternative approach to model update exchange for ML workloads: \emph{in-network aggregation}. In this approach, workers send their model updates over the network, where an aggregation primitive in the network sums the updates and distributes only the resulting value. Variations on this primitive have been proposed, over the years, for specialized supercomputer networks~\cite{bluegenel,bluegenep} and InfiniBand~\cite{sharp}. 
We demonstrate that it is possible to realize in-network aggregation in an Ethernet network and benefit ML applications.

\begin{figure}[t]
  \centering
  \includegraphics[width=0.45\textwidth]{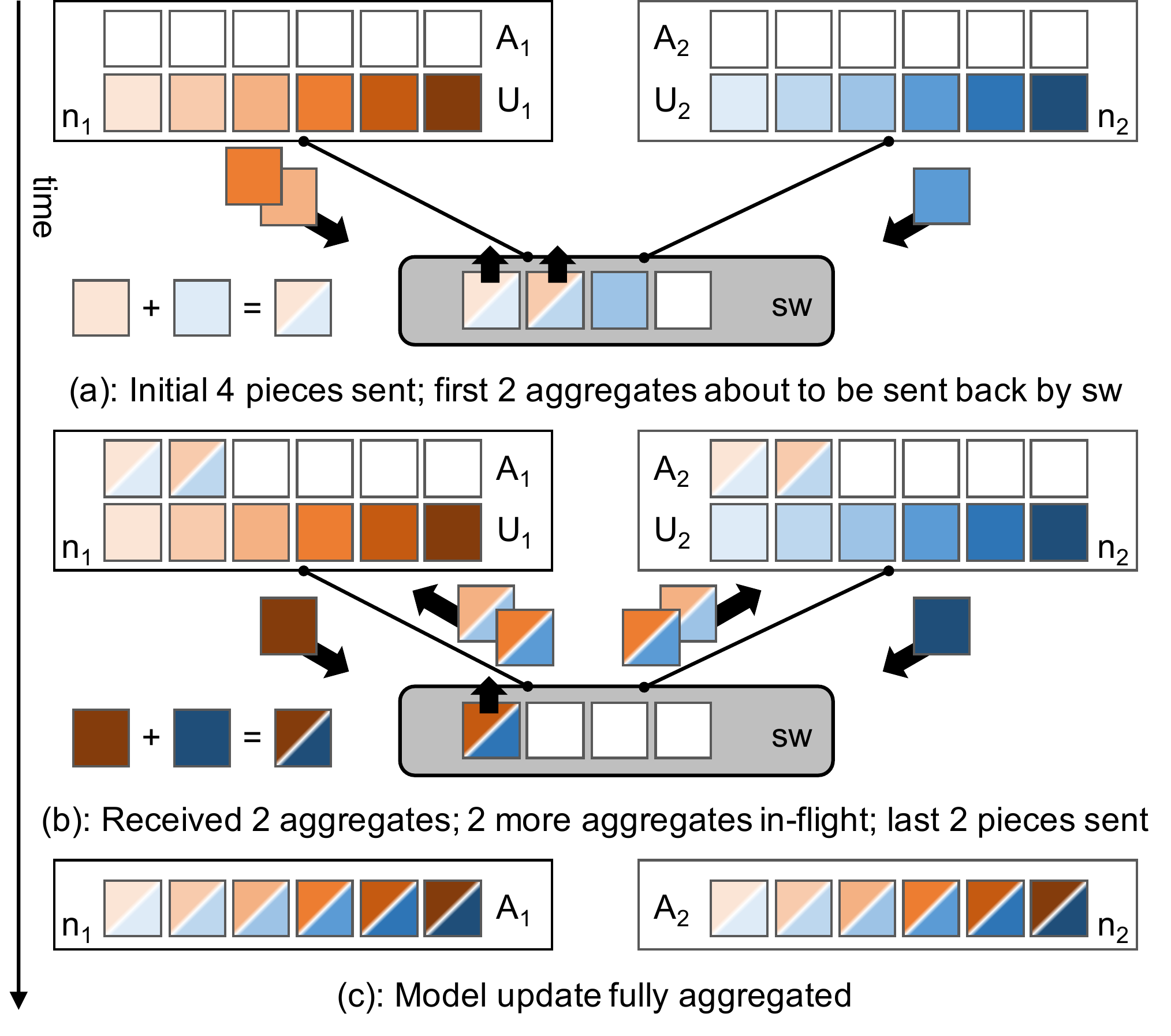}
  \caption{Example of in-network aggregation of model updates. $U_i$ is the model update computed by worker $i$. Workers stream pieces of model updates in a coordinated fashion. In the example, each workers can have at most 4 outstanding packets at any time to match the slots in the switch. The switch aggregates updates and multicasts back the values, which are collected into the aggregated model update $A_i$, then used to form the model parameters of the next iteration.}
  \label{fig:daiet}
\end{figure}

\smartparagraph{In-network aggregation offers a fundamental advantage} over both all-reduce and PS. It achieves the minimum possible latency and the minimum communication cost, quantified in data volume each worker sends and receives: $2|U|$ bytes, where $|U|$ is the total number of bytes to be aggregated. This is a significant improvement over the equivalent cost for bandwidth-optimal all-reduce, which is $4|U|\frac{n-1}{n}$~\cite{Patarasuk2009117}. 
The PS approach can match this communication cost of $2|U|$ bytes, at the expense of more resource cost; in the limit, it doubles the number of required machines and network bandwidth.\footnote{If the PS nodes are co-located with the worker nodes, then the effective bandwidth per node is halved, doubling latency.} Regardless of resource costs, in-network aggregation avoids end-host processing required to perform aggregation and, therefore, provides ``sub-RTT'' latency~\cite{netchain}, which the contrasted approaches cannot achieve.

\smartparagraph{Illustrating the advantages of in-network aggregation.} To characterize the extent to which communication is a bottleneck for training performance, we use our profile of eight DNN models from \S\ref{sec:network-bound}. We evaluate the impact of communication performance  using a trace of network-level events recorded during training. This trace captures real compute times and memory latency, including the latency for barrier events that precede each synchronization, but allows us to emulate different network speeds and computation patterns. In particular, our trace records the detailed timing of individual all-reduce invocations, so it faithfully accounts for potential overlap between communication and computation.\footnote{The ML toolkit adopts an optimization know as \emph{tensor fusion} or \emph{bucketing} that coalesces multiple all-reduce invocations to amortize setup overhead. Our traces reflect the effect of this optimization.}

We compare the performance of in-network aggregation (INA) with the current best practice, ring all-reduce (RAR).
Table~\ref{tab:synth} summarizes the batch processing speedup over the ring all-reduce performance. INA is consistently superior to RAR.
For communication-bound models (the four models in the 100 Gbps case), INA is up to 80\% and up to 67\% faster at 10 and 100 Gbps, respectively.
Note that this analysis reflects a \emph{theoretically optimal} implementation of RAR. The measured speedups (\S\ref{sec:eval}) of our real INA implementation are higher, because real RAR implementations do not achieve optimal performance; it is difficult to fully exploit all available bandwidth and avoid system overheads.

We also note that our profiling environment uses NVIDIA P100 devices. These are currently two-generation old GPU accelerators.
We investigate with real benchmarks in \S\ref{sec:eval} the impact of faster GPUs, which increases the relative impact of communication overheads.

\smartparagraph{Alternative: gradient compression.} Another way to reduce communication costs is to reduce the data volume of model updates using lossy compression. Proposed approaches include reducing the bit-width of gradient elements (quantization) or transmitting only a subset of elements (sparsification). These approaches come with tradeoffs: too much compression loss can impact the resulting model accuracy.

We adopt the results of a recent survey of gradient compression methods~\cite{grace} to emulate the behavior of Top-$k$~\cite{topk} and QSGD~\cite{alistarh2017qsgd} as two representative sparsification and quantization compressors, respectively. We use data from that study to identify the compression overhead and data reduction achieved. Our synthetic communication time, then, includes both the computational cost of compression and the communication cost of the all-gather operation used to exchange model updates (following their implementation~\cite{grace}).

We observe (Table~\ref{tab:synth}) that, although gradient compression decreases data volume, it is not necessarily superior to INA. In general, the computational cost of compression and decompression is non-negligible~\cite{grace, micro-fpga2018}; in some cases, it outweighs the communication-reduction benefits. In particular, INA outperforms QSGD on all workloads for both the 64 and 256 levels (6 and 8 bits). Similarly, Top-$k$ underperforms INA at the 10\% compression level, and even \emph{reduces} performance relative to RAR in the 100~Gbps setting. These observations agree with recent work~\cite{grace, micro-fpga2018}. In particular, Li et al.~\cite{micro-fpga2018} proposed additional hardware offloading, using a FPGA at every worker, to mitigate the cost of compression. As this requires additional hardware, our model does not consider it.

Gradient compression \emph{does} outperform INA when it can achieve high compression ratios, as with Top-$k$ at 1\%. However, in many cases, this level of compression either requires more training iterations to converge, or hurts the accuracy of the resulting model~\cite{grace}. For example, the NCF model achieves 95.8\% hit rate without compression after 20 epochs of training, while with Top-$k$ compression at 10\% it achieves 93.6\%. It fails to converge at 1\% compression. We report convergence comparisons for various models in Appendix \ref{app:compression}.

\begin{table}[t]
\resizebox{\columnwidth}{!}{%
\begin{tabular}{lrcccc}
 \toprule
Model     &  \textbf{INA} & \multicolumn{2}{c}{QSGD} &
                                                       \multicolumn{2}{c}{Top-$k$}   \\
  \cmidrule(lr){3-4}\cmidrule(lr){5-6}
& & 64 & 256 & 1\% & 10\%\\
\midrule
\multicolumn{6}{c}{10 Gbps}                                                   \\
\midrule
DeepLight  & 1.80 & 1.27 & 0.97     & \xmark9.24 (-1.1\%) & \cmark1.05 (-0.9\%)      \\
LSTM      & 1.77 & 1.27 & 0.97     & \xxmark7.49  & \xxmark1.05      \\
NCF       & 1.54 & 1.22 & 0.96     & \xxmark4.07 & \xmark1.05 (-2.2\%)      \\
BERT      & 1.54 & 1.20 & 0.98     & 3.45 (\dag) & 1.04 (\dag)      \\
VGG19     & 1.60 & 1.22 & 0.97     & \xmark2.13 (-10.4\%) & \xmark1.04 (-3.3\%)      \\
UGATIT    & 1.22 & 1.12 & 0.99     & \xxmark1.58 & \xxmark1.02      \\
ResNet-50 & 1.05 & 1.07 & 0.95     & \xmark1.15 (-1.7\%) & \cmark1.02 (+0.2\%)     \\
SSD       & 1.01 & 1.00 & 1.00     & \xmark1.01 (-2.4\%) & \cmark1.00 (-0.6\%)      \\ \midrule
\multicolumn{6}{c}{100 Gbps}                                           \\ \midrule
DeepLight & \smark1.67 & 0.93 & 0.78     & \xmark2.96 (-1.1\%) & \cmark0.47 (-0.9\%)      \\
LSTM       & \smark1.20 & 0.98 & 0.84     & \xxmark1.37 & \xxmark0.54     \\
NCF       & 1.22 & 1.00 & 0.85     & \xxmark1.22 & \xmark0.65 (-2.2\%)      \\
BERT      & \smark1.14 & 0.98 & 0.92     & ~1.27 (\dag) & 0.74 (\dag)\\
\bottomrule
\end{tabular}
}

{
\scriptsize
$^\dag$ The BERT task is fine-tuning from a pre-trained model. We are not able to assess the impact of compression on training.
}
  \caption{Analysis of batch processing speedup relative to ring
    all-reduce based on synthetic communication. For top-$k$
    compression, impact on model quality is shown in parentheses. 
    Accuracy penalties greater than 1\% are shaded in gray; red
    indicates failure to converge.
    At 100 Gbps, only the models that are network bottlenecked are shown.
\smark~indicates 100 Gbps cases where SwitchML achieves a higher batch processing speedup due to practical system overheads.}
  \label{tab:synth}
\end{table}

\section{Design}
\label{sec:design}

Our system, SwitchML, implements the aggregation primitive in a programmable dataplane switch. Such switches are now commercially available, with only a small cost premium compared to fixed-function switches~\cite{tofino}.
In-network aggregation is conceptually straightforward, but implementing it inside a programmable switch, however, is challenging. Although programmable switches allow placing computation into the network path, their limited computation and storage capabilities impose constraints on implementing gradient aggregation. The system must also tolerate packet loss, which, although uncommon in the cluster environment, is nevertheless possible for long-running DNN training jobs. {\system} addresses these challenges by appropriately dividing the functionality between the hosts and the switches, resulting in an efficient and reliable streaming aggregation protocol.

\subsection{Challenges}
\label{sec:chall}

\smartparagraph{Limited computation.} Mathematically, gradient aggregation is the average over a set of floating-point vectors. While a seemingly simple operation, it exceeds the capabilities of today's programmable switches. As they must maintain line rate processing, the number of operations they can perform on each packet is limited.  Further, the operations themselves can only be simple integer arithmetic/logic operations; neither floating-point nor integer division operations are possible. 

\smartparagraph{Limited storage.} Model updates are large. In each iteration, each worker may supply hundreds of megabytes of gradient values. This volume far exceeds the on-switch storage capacity, which is limited to a few tens of MB and must be shared with forwarding tables and other core switch functions. This limitation is unlikely to change in the future \cite{rmt}, given that speed considerations require dataplane-accessible storage to be implemented using on-die SRAM. 

\smartparagraph{Packet loss.} 
\system must be resilient to packet loss, without impact on efficiency or correctness (e.g., discarding part of an update or applying it twice because of retransmission).

\subsection{\system overview}

\system aims to alleviate communication bottlenecks for distributed ML training applications using in-network aggregation, in a practical cluster setting.\footnote{For simplicity, we assume dedicated bandwidth for the training jobs. We also assume that worker, link or switch failures are handled by the ML framework, as it is common in practice \cite{tensorflow, scaling-with-ps}.}
\system uses the following techniques to reduce communication costs while meeting the above challenges.

\smartparagraph{Combined switch-host architecture.} \system carefully partitions computation between end-hosts and switches to circumvent the restrictions of the limited computational power at switches. The switch performs integer aggregation, while end-hosts are responsible for managing reliability and performing more complex computations.

\smartparagraph{Pool-based streaming aggregation.} A complete model update far exceeds the storage capacity of a switch, so it cannot aggregate entire vectors at once. \system instead \emph{streams} aggregation through the switch: it processes the aggregation function on a limited number of vector elements at once. The abstraction that makes this possible is a pool of integer aggregators. In \system, end hosts handle the management of aggregators in a pool -- determining when they can be used, reused, or need more complex failure handling -- leaving the switch dataplane with a simple design.

\smartparagraph{Fault tolerant protocols.} 
We develop lightweight schemes to recover from packet loss with minimal overheads and adopt traditional mechanisms to solve worker or network failures.

\smartparagraph{Quantized integer-based aggregation.} Floating-point operations exceed the computational power of today's switches. We instead convert floating-point values to 32-bit integers using a block floating-point-like approach~\cite{bfp}, which is done efficiently at end hosts without impacting training accuracy.

We now describe each of these components in turn. To ease the presentation, we describe a version of the system in which packet losses do not occur.
We remove this restriction later. 

\subsection{Switch-side aggregation protocol}
\label{sec:switch-proto}

\begin{algorithm}[t!]
\footnotesize

\begin{distribalgo}[1]

   \INDENT{\textbf{Initialize State:}}
   \STATE n = number of workers
   \STATE pool[s], count[s] $:=$ $\{0\}$
   \ENDINDENT

   \INDENT{\textbf{upon receive} \mv{p(idx, off, vector)}}
   \STATE pool[\mv{p.idx}] $\gets$ pool[\mv{p.idx}] + \mv{p.vector} \COMMENT{+ is the vector addition}
   \STATE count[\mv{p.idx}]++
   \IF{count[\mv{p.idx}] = n}
      \STATE \mv{p.vector} $\gets$ pool[\mv{p.idx}]
      \STATE pool[\mv{p.idx}] $\gets$ 0; count[\mv{p.idx}] $\gets$ 0
      \MULTICAST{\mv{p}}
   \ELSE
      \DROP{\mv{p}}
   \ENDIF

  \ENDINDENT

\caption{Switch logic.}
\label{algo:switch_prog}
\end{distribalgo}
\end{algorithm}
We begin by describing the core network primitive provided by \system: in-switch integer aggregation. A \system switch provides a pool of $s$ integer aggregators, addressable by index. Each slot in the pool aggregates a vector of $k$ integers, which are delivered all at the same time in one update packet.
The aggregation function is the addition operator, which is commutative and associative -- meaning that the result does not depend on the order of packet arrivals. Note that addition is a simpler form of aggregation than ultimately desired: model updates need to be \emph{averaged}. As with the all-reduce approach, we leave the final division step to the end hosts, as the switch cannot efficiently perform this.

Algorithm~\ref{algo:switch_prog} illustrates the behavior of the aggregation primitive. A packet $p$ carries a pool index, identifying the particular aggregator to be used, and contains a vector of $k$ integers to be aggregated. Upon receiving a packet, the switch aggregates the packet's vector ($p.vector$) into the slot addressed by the packet's pool index ($idx$). Once the slot has aggregated vectors from each worker,\footnote{For simplicity, we show a simple counter to detect this condition. Later, we use a bitmap to track which workers have sent updates.} the switch outputs the result -- by rewriting the packet's vector with the aggregated value from that particular slot, and sending a copy of the packet to each worker. It then resets the slot's aggregated value and counter, releasing it immediately for reuse.

The pool-based design is optimized for the common scenario where model updates are larger than the memory capacity of a switch. It addresses two major limitations of programmable switch architectures. First, because switch memory is limited, it precludes the need to store an entire model update on a switch at once; instead, it aggregates pieces of the model in a streaming fashion. Second, it allows processing to be done at the packet level by performing the aggregation in small pieces, at most $k$ integers at a time. This is a more significant constraint than it may appear; to maintain a very high forwarding rate, today's programmable switches parse only up to a certain amount of bytes in each packet and allow computation over the parsed portion. Thus, the model-update vector and all other packet headers must fit within this limited budget, which is today on the order of a few hundred bytes; ASIC design constraints make it unlikely that this will increase dramatically in the future~\cite{rmt, drmt, domino}.
In our deployment, $k$ is 64 or 256.

\subsection{Worker-side aggregation protocol}

The switch-side logic above does not impose any constraints on which aggregator in the pool to use and when. Workers must carefully control which vectors they send to which pool index and, since pool size $s$ is limited, how they reuse slots.

There are two considerations in managing the pool of aggregators appropriately. For correctness, every worker must use the same slot for the same piece of the model update, and no slot can be simultaneously used for two different pieces. For performance, every worker must work on the same slot at roughly the same time to avoid long synchronization delays. To address these issues, we design a custom aggregation protocol running at the end hosts of ML workers.

For now, let us consider the non-failure case, where there is no packet loss. The aggregation procedure, illustrated in Algorithm~\ref{algo:worker_prog}, starts once every worker is ready to exchange its model update. Without loss of generality, we suppose that the model update's size is a multiple of $k$ and is larger than $k \cdot s$, where $k$ is the size of the vector aggregated in each slot and $s$ denotes the pool size.
Each worker initially sends $s$ packets containing the first $s$ pieces of the model update -- each piece being a contiguous array of $k$ values from offset $\textit{off}$ in that worker's model update $U$. Each of these initial packets is assigned sequentially to one of the $s$ aggregation slots. 

\begin{algorithm}[t!]
\footnotesize

\begin{distribalgo}[1]

   \FOR{i \textbf{in} 0 : s}
      \STATE \mv{p.idx} $\gets$ i
      \STATE \mv{p.off} $\gets$ k $\cdot$ i
      \STATE \mv{p.vector} $\gets$ U[\mv{p.off} : \mv{p.off} + k]
      \SEND{\mv{p}}
    \ENDFOR

   \REPEAT
      \RECV{\mv{p(idx, off, vector)}}
      \STATE A[\mv{p.off} : \mv{p.off}+k] $\gets$ \mv{p.vector}
      \STATE \mv{p.off} $\gets$ \mv{p.off} + k $\cdot$ s
      \IF{\mv{p.off} $<$ size(U)}
         \STATE \mv{p.vector} $\gets$ U[\mv{p.off} : \mv{p.off} + k]
         \SEND{\mv{p}}
      \ENDIF
   \UNTIL{A is incomplete}

\caption{Worker logic.}
\label{algo:worker_prog}
\end{distribalgo}
\end{algorithm}
After the initial batch of packets is sent, each worker awaits the aggregated results from the switch. Each packet received indicates that the switch has completed the aggregation of a particular slot. The worker consumes the result carried in the packet, copying that packet's vector into the aggregated model update $A$ at the offset carried in the packet ($p.\textit{off}$). The worker then sends a new packet with the \emph{next} piece of update to be aggregated. This reuses the same pool slot as the one just received, but contains a new set of $k$ parameters, determined by advancing the previous offset by $k \cdot s$.

A key advantage of this scheme is that it does not require any explicit coordination among workers and yet achieves agreement among them on which slots to use for which parameters. The coordination is implicit because the mapping between model updates, slots, and packets is deterministic. Also, since each packet carries the pool index and offset, the scheme is not influenced by reorderings. A simple checksum can be used to detect corruption and discard corrupted packets.

This communication scheme is self-clocked after the initial $s$ packets. This is because a slot cannot be reused until all workers have sent their contribution for the parameter update for the slot. When a slot is completed, the packets from the switch to the workers serve as flow-control acknowledgments that the switch is ready to reuse the slot, and the workers are free to send another packet. Workers are synchronized based on the rate at which the system aggregates model updates. The pool size $s$ determines the number of concurrent in-flight aggregations; as we elaborate in Appendix~\S\ref{sec:tuning}, the system achieves peak bandwidth utilization when $k \cdot s$ (more precisely, $b \cdot s$ where $b$ is the packet size -- 1100 bytes in our setting) exceeds the bandwidth-delay product of the inter-server links.

\subsection{Dealing with packet loss}
\label{sec:packet-loss}

Thus far, we have assumed packets are never lost. Of course, packet loss can happen due to either corruption or network congestion. With the previous algorithm, even a single packet loss would halt the system. A packet loss on the ``upward'' path from workers to the switch prevents the switch from completing the corresponding parameter aggregations.  The loss of one of the result packets that are multicast on the ``downward'' paths not only prevents a worker from learning the result but also prevents it from ever completing $A$.

We tolerate packet loss by retransmitting lost packets. In order to keep switch dataplane complexity low, packet loss detection is done by the workers if they do not receive a response packet from the switch in a timely manner. However, na\"ive retransmission creates its own problems.  If a worker retransmits a packet that was actually delivered to the switch, it can cause a model update to be applied twice to the aggregator. On the other hand, if a worker retransmits a packet for a slot that was actually already fully aggregated (e.g., because the response was lost), the model update can be applied to the wrong data because the slot could have already been reused by other workers who received the response correctly.
Thus, the challenges are (1) to be able to differentiate packets that are lost on the upward paths versus the downward ones; and (2) to be able to retransmit an aggregated response that is lost on the way back to a worker.

\begin{algorithm}[t!]
\footnotesize

\begin{distribalgo}[1]

   \INDENT{\textbf{Initialize State:}}
   \STATE n = number of workers
   \STATE pool[2, s], count[2, s], seen[2, s, n] $:=$ $\{0\}$
   \ENDINDENT

   \INDENT{\textbf{upon receive} \mv{p(wid, ver, idx, off, vector)}}
   \IF{seen[\mv{p.ver, p.idx, p.wid}] = 0}
      \STATE seen[\mv{p.ver, p.idx, p.wid}] $\gets$ 1
      \STATE seen[\mv{(p.ver+1)\%2, p.idx, p.wid}] $\gets$ 0
      \STATE count[\mv{p.ver, p.idx}] $\gets$ (count[\mv{p.ver, p.idx}]+1)\%n
      \IF{count[\mv{p.ver, p.idx}] = 1}
         \STATE pool[\mv{p.ver, p.idx}] $\gets$ \mv{p.vector}
      \ELSE
         \STATE pool[\mv{p.ver, p.idx}] $\gets$ pool[\mv{p.ver, p.idx}] + \mv{p.vector}
      \ENDIF
      \IF{count[\mv{p.ver, p.idx}] = 0}
         \STATE \mv{p.vector} $\gets$ pool[\mv{p.ver, p.idx}]
         \MULTICAST{\mv{p}}
      \ELSE
         \DROP{\mv{p}}
      \ENDIF
   \ELSE
      \IF{count[\mv{p.ver, p.idx}] = 0}
         \STATE \mv{p.vector} $\gets$ pool[\mv{p.ver, p.idx}]
         \FORWARD{\mv{p} to \mv{p.wid}}
      \ELSE
         \DROP{\mv{p}}
      \ENDIF
   \ENDIF

  \ENDINDENT

\caption{Switch logic with packet loss recovery.}
\label{algo:switch_prog_loss}
\end{distribalgo}
\end{algorithm}
We modify the algorithms to address these issues by keeping two additional pieces of switch state. First, we explicitly maintain information as to which workers have already contributed updates to a given slot. This makes it possible to ignore duplicate transmissions. Second, we maintain a \emph{shadow copy} of the \emph{previous} result for each slot. That is, we have two copies or versions of each slot, organized in two pools; workers alternate between these two copies to aggregate successive chunks that are assigned to the same slot. The shadow copy allows the switch to retransmit a dropped result packet for a slot even when the switch has started reusing the slot for the next chunk.

The key insight behind this approach's correctness is that, even in the presence of packet losses, our self-clocking strategy ensures that no worker node can ever lag \emph{more than one chunk} behind any of the others for a particular slot.
This invariant is because the switch will not release a slot to be reused, by sending a response, until it has received an update packet from \emph{every} worker for that slot. Furthermore, a worker will not send the next chunk for a slot until it has received the response packet for the slot's previous chunk, preventing the system from moving ahead further. As a result, it is sufficient to keep only one shadow copy.

Besides obviating the need for more than one shadow copy, this has a secondary benefit: the switch does not need to track full phase numbers (or offsets); a single bit is enough to distinguish the two active phases for any slot.

\begin{algorithm}[t!]
\footnotesize

\begin{distribalgo}[1]

   \FOR{i \textbf{in} 0 : s}
      \STATE \mv{p.wid} $\gets$ Worker ID
      \STATE \mv{p.ver} $\gets$ 0
      \STATE \mv{p.idx} $\gets$ i
      \STATE \mv{p.off} $\gets$ k $\cdot$ i
      \STATE \mv{p.vector} $\gets$ U[\mv{p.off} : \mv{p.off} + k]
      \SEND{\mv{p}}
      \STATE start\_timer(\mv{p})
   \ENDFOR

   \REPEAT
      \RECV{\mv{p(wid, ver, idx, off, vector)}}
      \STATE cancel\_timer(\mv{p})
      \STATE A[\mv{p.off} : \mv{p.off}+k] $\gets$ \mv{p.vector}
      \STATE \mv{p.off} $\gets$ \mv{p.off} + k $\cdot$ s
      \IF{\mv{p.off} $<$ size(U)}
         \STATE \mv{p.ver} $\gets$ (\mv{p.ver}+1)\%2
         \STATE \mv{p.vector} $\gets$ U[\mv{p.off} : \mv{p.off} + k]
         \SEND{\mv{p}}
         \STATE start\_timer(\mv{p})
      \ENDIF
   \UNTIL{A is incomplete}

   \BLANK
   \INDENT{\textbf{upon timeout} \mv{p} \textit{/* Timeout Handler */}}
      \SEND{\mv{p}}
      \STATE start\_timer(\mv{p})
   \ENDINDENT

\caption{Worker logic with packet loss recovery.}
\label{algo:worker_prog_loss}
\end{distribalgo}
\end{algorithm}
In keeping with our principle of leaving protocol complexity to end hosts, the shadow copies are kept in the switch but managed entirely by the workers. The switch simply exposes the two pools to the workers, and the packets specify which slot acts as the active copy and which as the shadow copy by indicating a single-bit pool version (\textit{ver}) field in each update packet. The pool version starts at 0 and alternates each time a slot with the same \textit{idx} is reused.

Algorithms~\ref{algo:switch_prog_loss} and \ref{algo:worker_prog_loss} show the details of how this is done. An example illustration is in Appendix \ref{app:example}. In the common case, when no losses occur, the switch receives updates for slot $idx$, pool \textit{ver} from all workers. 
When workers receive the response packet from the switch, they change the pool by flipping the \textit{ver} field -- making the old copy the shadow copy -- and send the next phase updates to the other pool.

A timeout detects packet loss at each worker. When this occurs, the worker does not know whether the switch received its previous packet or not. Regardless, it retransmits its earlier update with the same slot $idx$ and $ver$ as before. This slot is guaranteed to contain the state for the same aggregation in the switch. The \textit{seen} bitmask indicates whether the update has already been applied to the slot. If the aggregation is already complete for a slot, and the switch yet receives an update packet for the slot, the switch recognizes the packet as a retransmitted packet and replies with a unicast packet containing the result. 
The result in one slot is overwritten for reuse only when there is the certainty that all the workers have received the slot's aggregated result. Slot reuse happens when all the workers have sent their updates to the same slot of the other pool, signaling that they have all moved forward. Note this scheme works because the completion of aggregation for a slot $idx$ in one pool \textit{safely and unambiguously} confirms that the previous aggregation result in the shadow copy of slot $idx$ has indeed been received by every worker.

This mechanism's main cost is switch memory usage: keeping a shadow copy doubles the memory requirement, and tracking the \textit{seen} bitmask adds additional cost. This may appear problematic, as on-switch memory is a scarce resource. In practice, however, the total number of slots needed -- tuned based on the network bandwidth-delay product (Appendix \S\ref{sec:tuning}) -- is much smaller than the switch's memory capacity.

\subsection{Dealing with floating-point numbers}

DNN training commonly uses floating-point numbers, but current programmable switches do not natively support them. We explored two approaches to bridging this gap.

Floating-point numbers are already an approximation. SGD and similar algorithms are defined over real numbers. Floating-point numbers approximate real numbers by trading off range, precision, and computational overhead to provide a numerical representation that can be broadly applied to applications with widely different properties. However, many other approximations are possible. An approximation designed for a specific application can obtain acceptable accuracy with lower overhead than standard floating-point offers. 

In recent years, the community has explored many specialized numerical representations for DNNs. These representations exploit the properties of the DNN application domain to reduce the cost of communication and computation. For instance, NVIDIA Volta and Ampere GPUs~\cite{volta, nvidia-a100} include mixed-precision (16-/32-bit) TPUs that can train with accuracy matching full-precision approaches. Other work has focused on gradient exchange for SGD, using fixed-point quantization, dithering, or sparsification to reduce both the number of bits and the gradient elements transmitted~\cite{lin2018deep, wen2017terngrad, DIANA, seide2014, bernstein2018signsgd, pmlr-v80-bernstein18a,dorefa}. Further, others have explored block floating-point representations \cite{bfp, flexpoint}, where a single exponent is shared by multiple tensor elements, reducing the amount of computation required to perform tensor operations. This innovation will continue; our goal is not to propose new representations but to demonstrate that techniques like those in the literature are practical with programmable switches.

We use a numeric representation, inspired by block floating-point, that combines 32-bit fixed-point addition in the switch with adaptive scaling on the workers. This representation is used only when aggregating gradients; all other data (weights, activations) remain in 32-bit floating-point representation. 

To implement our representation, we scale gradient values using a per-packet scaling factor $f$, which is automatically determined for each use of an aggregator slot in the switch. The scaling factor is set so that the maximum aggregated floating point value within a block of gradients is still representable as a 32-bit fixed point value. 
Namely, let $h$ be the largest absolute value of a block of gradients; $f$ is set to $(2^{31} - 1) / (n \cdot 2^{m})$, where $m$ is the exponent of $h$ rounded up to a power of 2 and $n$ is the number of workers. Appendix~\ref{app:quant} formally analyzes the precision of this representation.

To realize this quantization of floating-point values, workers need to agree on a global value of $m$ prior to sending the corresponding block of gradients. We devise a simple look-ahead strategy: when workers send the $j$-th block to slot $i$, they include their local block $j+1$'s maximum gradient (rounded up to a power of 2). The switch identifies the global maximum $m$ and piggy-backs that value when sending the aggregated gradients of the $j$-th block.

We verify experimentally that this communication quantization allows training to similar accuracy in a similar number of iterations as an unquantized network.
We illustrate the convergence behavior by training a ResNet110 model on CIFAR10 dataset for 64,000 steps (about 41 epochs) using 8 workers.
Figure~\ref{fig:resnet110} shows the test accuracy over time.
The accuracy obtained by SwitchML (about 91-93\% in the last 5 points) is similar to that obtained by training with TensorFlow on the same
worker setup, and it matches prior results~\cite{resnet} with the same
hyperparameter settings. The training loss curves (not shown) show the same similarity.
In Appendix \ref{app:quant}, we further give a detailed convergence analysis for the aforementioned representation on models in Table \ref{tab:dnns}.

\begin{figure}[t]
 \centering
 \includegraphics[width=0.45\textwidth]{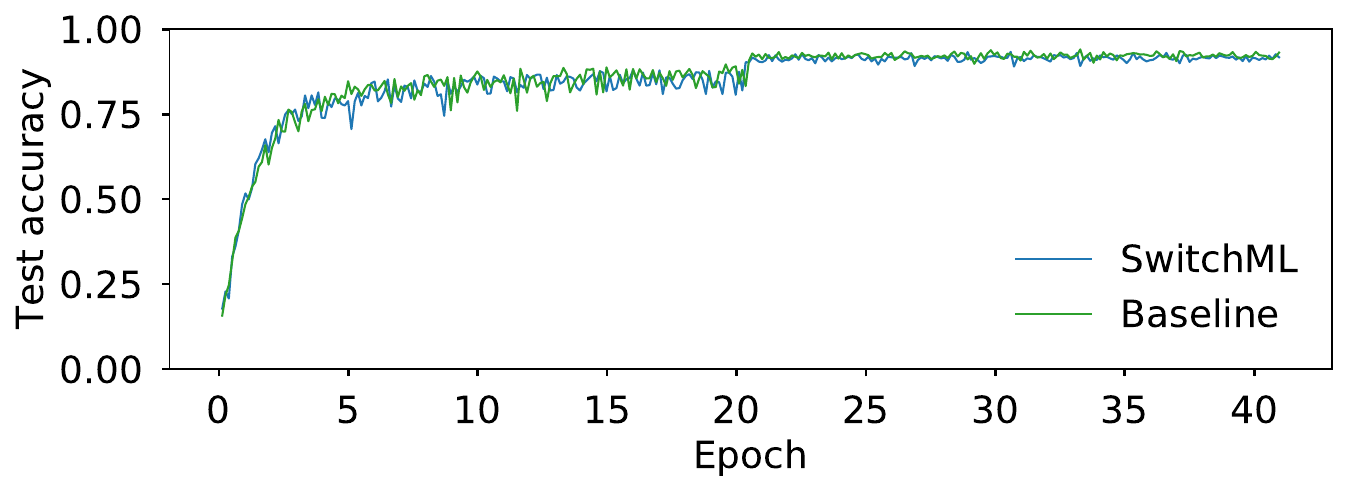}
 \caption{Test accuracy of ResNet110 on CIFAR10. SwitchML achieves similar accuracy to the baseline.}
 \label{fig:resnet110}
\end{figure}

While the above representation is used in the remainder of the paper, we also explored the implementation of a restricted form of 16-bit floating-point. In this version, the switch converts each 16-bit floating-point value in the incoming packets into a fixed-point value and then performs aggregation. When generating responses, the switch converts fixed-point values back into floating-point values. Due to resource limitations in the switch, we were only able to support half the dynamic range of the 16-bit floating-point format; we expect this to lead to poor convergence during training. Conversely, our 32-bit integer format uses minimal switch resources,  provides good dynamic range, and has a minimal overhead on workers.
A 16-bit format would provide a bandwidth benefit (\S\ref{sec:scalingandtypeconverstion}).

\section{Implementation}

We build \system as a collective library which we integrate in PyTorch's DistributedDataParallel module and in TersorFlow via Horovod~\cite{horovod}. 
\system is implemented as a
worker component written as $\sim$3,100 LoCs in C++ and a switch
component realized in P4~\cite{p4} with $\sim$3,700 LoCs. The worker is built atop Intel DPDK. We have also built a RDMA-capable
implementation, but it is not yet integrated with the training
frameworks.
Here, we highlight a few salient aspects of our implementation. Appendix \ref{app:impl} describes more details.

Our P4 program distributes
aggregation across multiple stages of the ingress pipeline, and also
implements flow control, retransmission, and exponent-calculation
logic. It uses the traffic manager subsystem to send multiple copies
of result packets. It can process 64 elements per packet using one
switch pipeline, and 256-element (1024-byte) packets using all four
switch pipelines.
On the worker side, we process each packet in a run-to-completion
fashion and scale to multiple CPU cores using DPDK and Flow Director. We use up to 8 cores per worker. This scales well because we shard slots and chunks of tensors across cores without any shared state.
The ML framework invokes our synchronous API whenever model updates are ready. In practice, model updates consist of a set of tensors that are aggregated independently but sequentially.

\smartparagraph{Supporting large packets.}
\label{sec:largerpackets}
Good bandwidth efficiency requires processing enough integer elements in each packet to offset the network framing overhead.
Our P4 program can parse and aggregate 64 $\times$
4-byte elements per packet, but can only read 32 elements per packet
when aggregation is complete. With framing overheads, a
32-element payload would limit goodput to 63\% of line rate. Our P4 program supports larger packets in two
additional configurations for better efficiency: a 64-element
configuration with 77\% goodput, and a 256-element one with 93\%
goodput.

We support larger packets through {\it recirculation}: sending packets
through the switch pipelines multiple times. Our 64-element design
uses a single pipeline. It makes one additional pass through the
pipeline \emph{only} when an output packet is broadcast in order to read the results: this separation of reads and writes allows us to write 64 elements in a single pass. The internal recirculation ports provided by the
chip provide sufficient bandwidth. To support 256 elements, we
recirculate packets through all four switch pipelines. This requires
placing switch ports into loopback mode for more recirculation
bandwidth, leaving 16 $\times$ 100 Gbps bandwidth available for workers.
When a slot is complete, we recirculate again through all the pipelines to read the results. Tofino has sufficient bandwidth to do this recirculation at 1.6 Tbps, and the latency scales deterministically with the number of pipeline passes: we measure an additional 520 ns per pass.

\smartparagraph{Supporting RDMA.}
\label{sec:rdma}
Our host-side framework, even using DPDK and
multiple cores, has difficulty achieving 100 Gbps throughput due to
packet processing costs. We address this by implementing a subset of
RDMA in the switch. This allows workers to offload packet processing:
the RDMA NIC breaks large messages into individual packets.
Specifically, we use RoCE~v2~\cite{rocev2} in Unreliable
Connected (UC) mode~\cite{rdmaaware}. This mode, which does not
require any of RoCE's link-level flow control mechanisms, supports
multi-packet messages and detects packet drops, but does not implement
retransmission. \system continues to rely on its existing reliability
mechanism. Timeouts and duplicate packets are handled as before,
except that a timeout forces a client to retransmit the entire
multi-packet message. To balance the benefit of offload with the cost of retransmission, we use small, multi-packet messages (generally 16 packets per message). Although retransmissions are more expensive, the common case is much faster, even though we use a single CPU core.

RDMA Write Immediate messages are used for all communication, allowing
data to move directly between the switch and GPUs, with client CPUs
handling protocol operations. \system metadata is encoded in RDMA
headers. Concurrent messages are sent on separate queue pairs to allow
packets to interleave; queue pair IDs and access keys are negotiated
with the switch control plane during job setup.
The switch sends aggregated results
by generating RDMA Write messages to the destination buffer.

\section{Evaluation}
\label{sec:eval}

We analyze the performance benefits of \system using standard benchmarks on popular models in TensorFlow and PyTorch and using microbenchmarks to compare it to state-of-the-art collective communications libraries and PS scenarios.

\smartparagraph{Testbed.}
We conduct most of our experiments on a testbed of 8 machines, each with 1 NVIDIA P100 16 GB GPU, dual 10-core CPU Intel Xeon E5-2630v4 at 2.20 GHz, 128 GB of RAM, and 3 $\times$ 1 TB disks for storage (as single RAID).
To demonstrate scalability with 16 nodes, we further use 8 machines
with dual 8-core CPU Intel Xeon Silver 4108 at 1.80 GHz. 
Moreover, we use a Wedge100BF-65X programmable switch with Barefoot Networks' Tofino chip~\cite{tofino}.
Every node is networked at both 10 and 100 Gbps.

\smartparagraph{Performance metrics.}
We mostly focus on two performance metrics.
We define \emph{tensor aggregation time} (TAT) as the time to aggregate a tensor starting from the time a worker is ready to send it till the time that worker receives the aggregated tensor; lower is better.
We also report \emph{aggregated tensor elements} (ATE) per unit of time, for presentation clarity; higher is better.
For these metrics, we collect measurements at each worker for aggregating 100 tensors of the same size, after 10 warmups.
We measure \emph{training throughput} defined in terms of the numbers of training samples processed per second.
We measure throughput for 100 iterations that follow 100 warmups.
A variant of training throughput is the batch processing throughput, which we use to analyze performance by replaying profile traces. This throughput metric includes communication and computation costs, but excludes the time to load data.

\smartparagraph{Benchmarks.}
We evaluate \system by training with 8 DNNs introduced in Table~\ref{tab:dnns}. The detailed configuration of the benchmarks is in Table~\ref{tab:dnns-config} in Appendix~\ref{app:impl}.
Half of the benchmarks execute on PyTorch and half on TensorFlow.

\smartparagraph{Setup.}
As a baseline, we run both PyTorch with native distributed data-parallel module and TensorFlow with Horovod. By default, we use NCCL as the communication library, and use both TCP and RDMA as the transport protocol.
Our default setup is to run experiments on 8 workers.

\subsection{Tensor aggregation microbenchmarks}
\label{sec:micro}

To illustrate \system's efficiency in comparison to other communication strategies, we devise a communication-only microbenchmark that performs continuous tensor aggregations, without any gradient computation on the GPU. We verify that the tensors -- initially, all ones -- are aggregated correctly.
We test with various tensor sizes from 50 MB to 1.5 GB. We observe that the number of aggregated tensor elements per time unit (ATE/s) is not very sensitive to the tensor size.
Thus, we report results for 100 MB tensors only.

For these experiments, we benchmark \system against the popular all-reduce communication libraries (Gloo~\cite{gloo} and NCCL~\cite{nvidia-nccl}). We further compare against a parameter server-like scenario, i.e., a set of worker-based processes that assist with the aggregation. To this end, we build a DPDK-based program that implements streaming aggregation as in Algorithm \ref{algo:switch_prog}.
To capture the range of possible PS performance, we consider two scenarios: (1) when the PS processes run on dedicated machines, effectively doubling the cluster size, and (2) when a PS process is co-located with every worker.
We choose to run as many PS processes (each using 8 cores) as workers so that tensor aggregation workload is equally spread among all machines (uniformly sharded) and avoids introducing an obvious performance bottleneck due to oversubscribed bandwidth, which is the case when the ratio of workers to PS nodes is greater than one. 

\begin{figure}[t]
 \centering
 \includegraphics[width=0.48\textwidth]{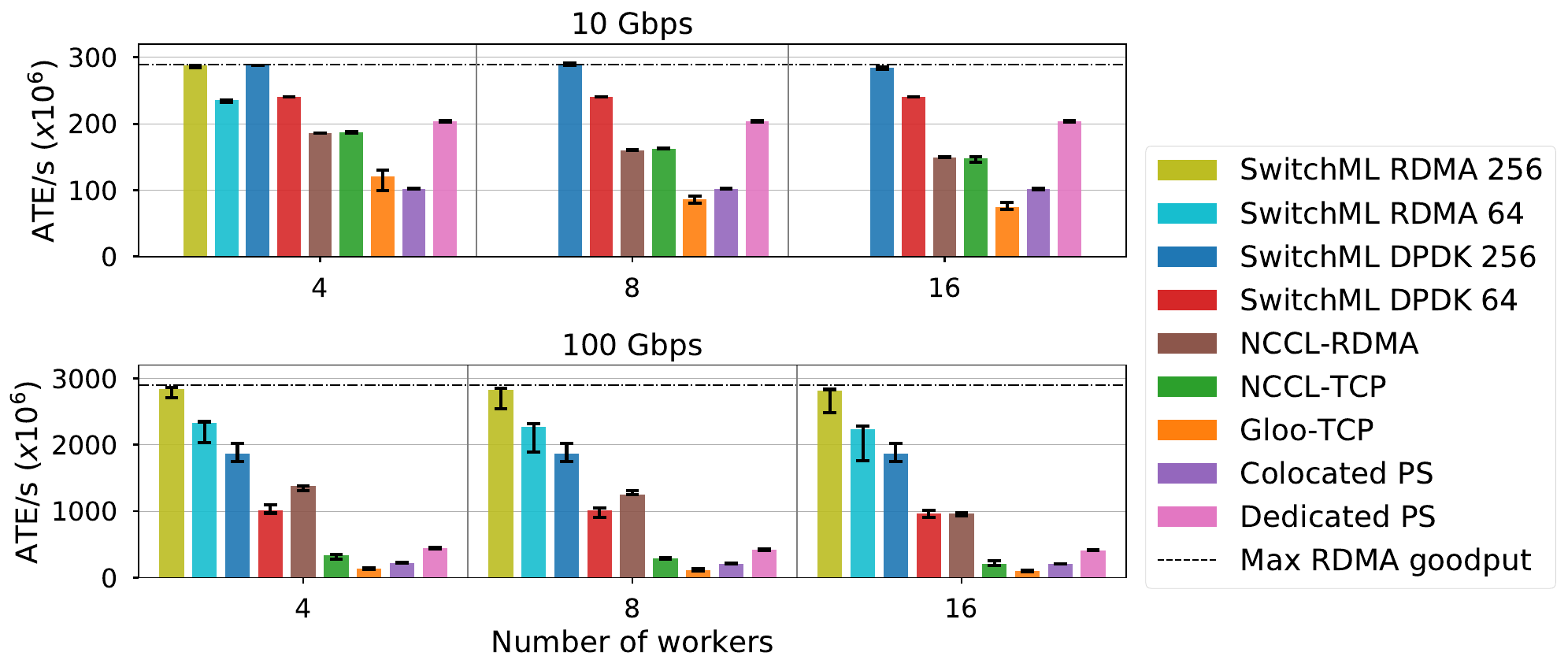}
 \caption{Microbenchmarks showing aggregated tensor elements per second on a 10 (top) and 100 (bottom) Gbps network as workers increase.}
 \label{fig:microbenchmark}
\end{figure}

Figure \ref{fig:microbenchmark} shows the results at 10 and 100 Gbps on three cluster sizes.
The results demonstrate the efficiency of \system: its highest-performing variant, which uses RDMA with 256-value (1024-byte payload) packets, is within 2\% of the maximum achievable goodput. Using smaller packets ($k=64$ instead of $256$) has a noticeable performance impact, underscoring the importance of our multi-pipeline design. The DPDK implementation has additional host-side overhead that prevents it from achieving full link utilization at 100~Gbps. In spite of this, \system can still outperform the best current all-reduce system, NCCL, even when it uses RDMA and \system does not.
Moreover, \system always maintains a predictable rate of ATE/s regardless of the number of workers.
This trend should continue with larger clusters.

The Dedicated PS approach (with 256 values per packet) -- while using \emph{twice} the number of machines \emph{and} network capacity -- falls short of matching \system DPDK performance.
Unsurprisingly, using the same number of machines as \system, the Colocated PS approach reaches only half of Dedicated PS performance.
Our PS implementation is simpler than (and should outperform) a traditional PS, as we do not store the entire model in memory. 
It demonstrates that, in principle, our aggregation protocol could be run entirely in software on a middlebox, but with lower performance: in-network aggregation inherently requires fewer resources than host-based aggregation.

\begin{figure}[t]
 \centering
 \includegraphics[width=0.48\textwidth]{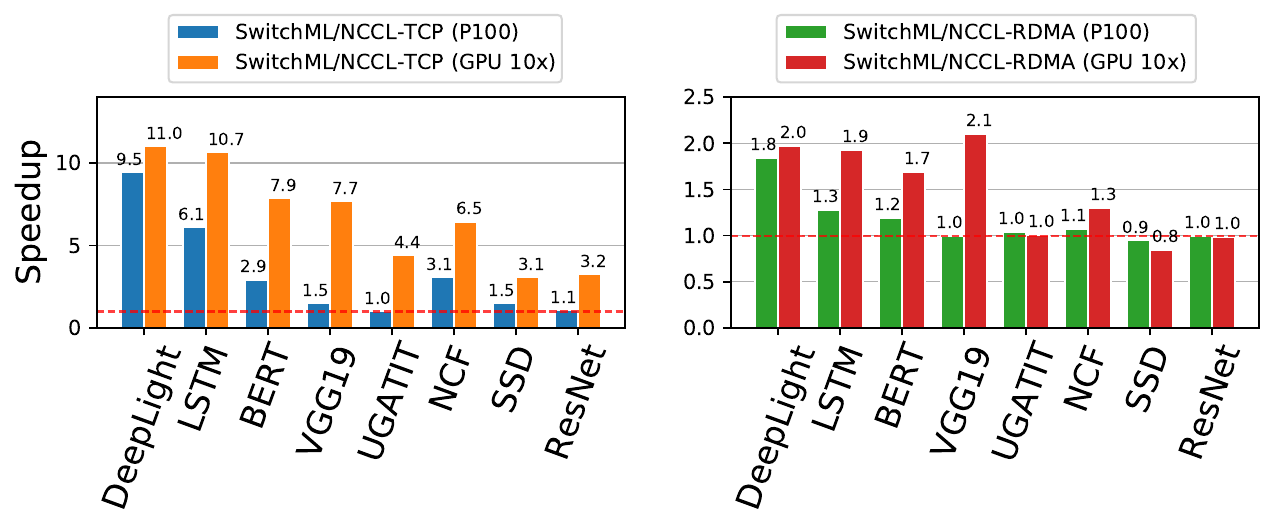}
 \caption{Training batch processing speedup at 100 Gbps considering a P100 GPU and a 10$\times$ faster GPU.}
 \label{fig:trace_speedup}
\end{figure}

\begin{figure}[t]
 \centering
 \includegraphics[width=0.48\textwidth]{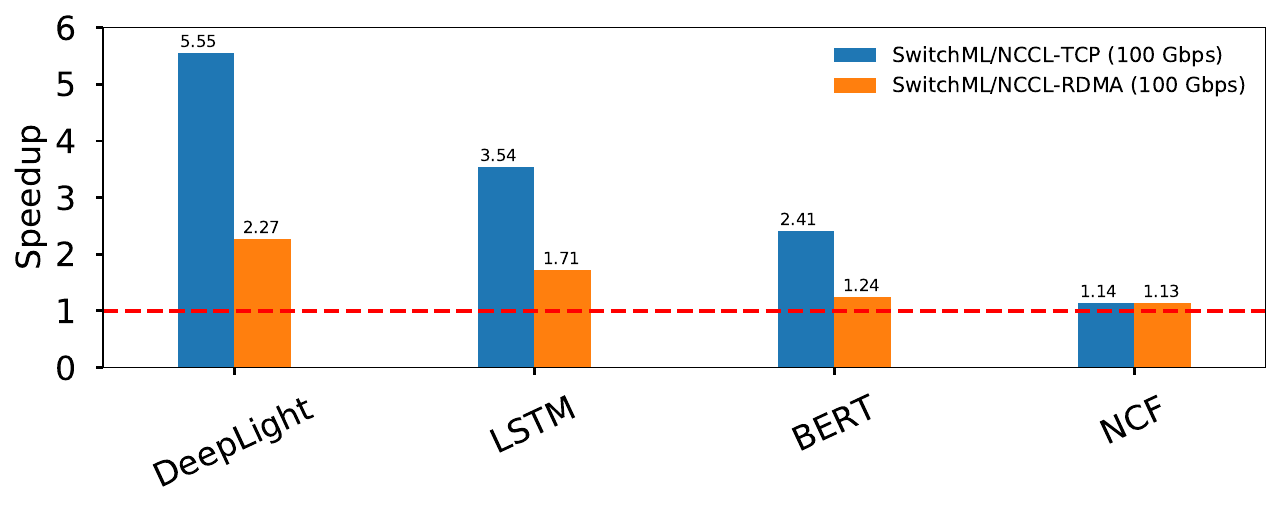}
 \caption{Training performance speedup normalized to NCCL for varying transport protocols.}
 \label{fig:training_speedup}
\end{figure}

\subsection{\system improves training speed}
\label{sec:eval:training}

We now analyze training performance on eight popular DNN benchmarks.
We normalize results to NCCL as the underlying communication backend of PyTorch and TensorFlow.

Figure \ref{fig:trace_speedup} reports the speedup for processing a training batch for SwitchML compared to NCCL at 100 Gbps. These results replay the profile traces collected on our cluster (\S\ref{sec:network-bound}), allowing us to report both the speedup on our testbed GPUs (P100s, which are two generations old) and hypothetical GPUs that are 10$\times$ faster (by uniformly scaling the traces by this factor).
This scaling factor is roughly comparable to the fastest GPUs available today (the A100 benchmarks at 4.2$\times$ the V100~\cite{nvidia-a100-4x}, which in turn is 1.4-2.2$\times$ faster than our P100~\cite{dell-v100-speedup}).  This emulation lets us evaluate the setting where a fast network is paired with fast GPUs.
\system uses the DPDK implementation with 256-value packets.

As expected, \system accelerates batch processing especially for the larger DNNs (up to one order of magnitude when considering NCCL-TCP).
The speedup over NCCL-RDMA is at most 2.1$\times$, which is in line with the fundamental 2$\times$ advantage of INA over RAR (\S\ref{sec:ina}). In most cases, the measured speedup is higher than the emulated communication results (Table \ref{tab:synth}) predict, because NCCL's RAR implementation does not achieve the theoretical maximum efficiency. 

\system provides significant benefits for many, but not all, real-world DNNs, even with 100 Gbps networks. For example, DeepLight and LSTM enjoy major improvements. BERT sees a somewhat lower speedup, in part because its gradient consists of many relatively small ($\sim$60 MB) tensors. Similarly, NCF, a relatively small model, has a modest speedup. Other models, like UGATIT, SSD, and ResNet are simply not network-bound at 100 Gbps. SSD is a particularly challenging case: not only is it a small model that would require an $\alpha=15.2\times$ faster GPU to become network-bound (Table \ref{tab:dnns}), it also makes many aggregation invocations for small gradients. The overheads of starting an aggregation are not well amortized, especially in the 10$\times$ scaled scenario.

Finally, we consider the end-to-end speedup on a complete training run with 16 workers. We focus on the four models that are network-bottlenecked at 100 Gbps.
Figure \ref{fig:training_speedup} shows the training performance speedup compared to NCCL using RDMA and TCP. These measurements use \system's DPDK implementation, with 256-value packets; we expect a larger speedup once \system's RDMA implementation is integrated with the training framework. Even so, \system's speedups range between 1.13-2.27$\times$ over NCCL-RDMA and 2.05-5.55$\times$ over NCCL-TCP. 
The results are not directly comparable to Figure~\ref{fig:trace_speedup}, because (1) they use a larger 16-node cluster, and (2) they report total end-to-end iteration time, which also includes data loading time. Our deployment does not use any optimized techniques for data loading, an orthogonal problem being addressed by other work (e.g., DALI~\cite{nvidia-dali}).

\subsection{Overheads}
\label{sec:overheads}

\smartparagraph{Packet loss recovery.}
We study how packet loss affects TAT.
To quantify the change in TAT due to packet loss, we experiment with a uniform random loss probability between 0.01\% and 1\% applied on every link.
The retransmission timeout is set to 1 ms.
We run microbenchmark experiments in similar scenarios as \S\ref{sec:micro}. We report a few representative runs.

Figure \ref{fig:timer_overhead} measures the inflation in TAT with different loss probabilities. \system completes tensor aggregation significantly faster than Gloo or NCCL when the loss is 0.1\% or higher. A loss probability of 0.01\% minimally affects TAT in either case.
To better illustrate the behavior of \system, we show in Figure \ref{fig:load_over_time} the evolution of packets sent per 10 ms at a representative worker for 0.01\% and 1\% loss. We observe that \system generally maintains a high sending rate -- relatively close to the ideal rate -- and quickly recovers by retransmitting dropped packets.
The slowdown past the 150 ms mark with 1\% loss occurs because some slots are unevenly affected by random losses and \system does not apply any form of work-stealing to rebalance the load among aggregators. This presents a further opportunity for optimization. 

\smartparagraph{Tensor scaling and type conversion.}
\label{sec:scalingandtypeconverstion}
We analyze whether any performance overheads arise due to the tensor scaling operations (i.e., multiply updates by $f$ and divide aggregates by~$f$) and the necessary data type conversions: float32-to-int32 $\rightarrow$ htonl $\rightarrow$ ntohl $\rightarrow$ int32-to-float32.

To quantify overheads, we use int32 as the native data type while running the microbenchmarks. This emulates a native float32 scenario with no scaling and conversion operations.
We also illustrate the potential improvement of quantization to single-precision (float16) tensors, which halves the volume of data to be sent to the network. (We include a conversion from/to float32.)
This setting is enabled by the ability to perform at line rate, in-switch type conversion (float16 $\leftrightarrow$ int32), which we verified with the switch chip vendor. However, for this experiment, we emulate this by halving the tensor size.

\begin{figure}[t]
 \centering
 \includegraphics[width=0.40\textwidth]{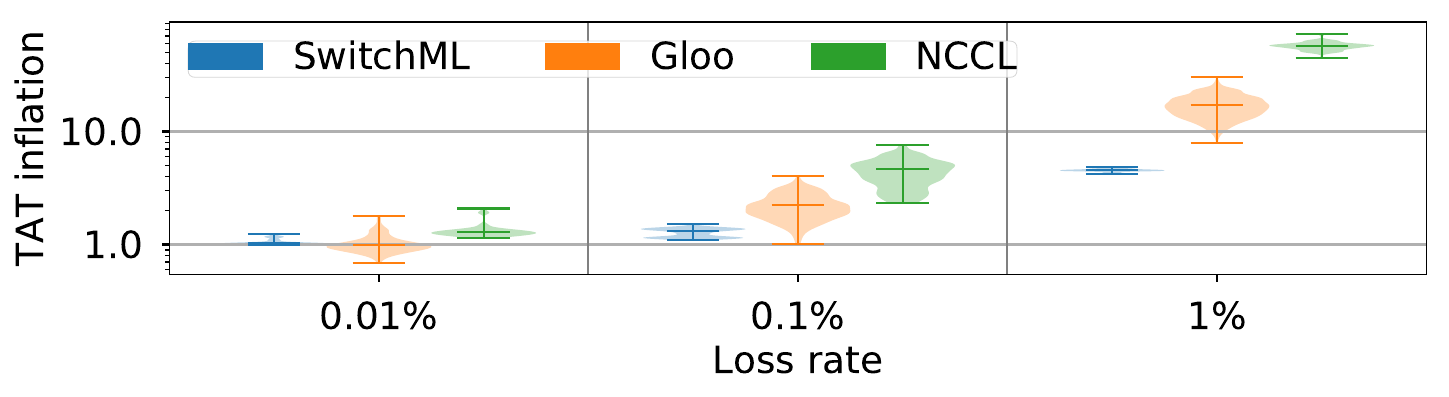}
 \caption{Inflation of TAT due to packet loss and recovery. Results are normalized to a baseline scenario where no loss occurs and the worker implementation does not incur any timer-management overhead.}
 \label{fig:timer_overhead}
\end{figure}

\begin{figure}[t]
 \centering
 \includegraphics[width=0.40\textwidth]{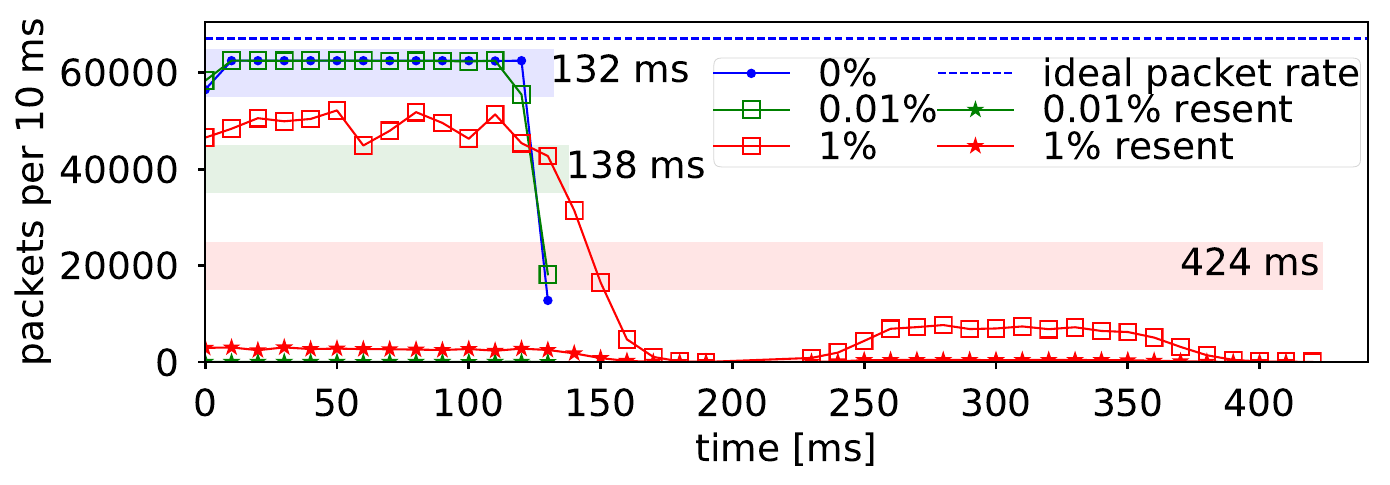}
 \caption{Timeline of packets sent per 10 ms during an  aggregation with 0\%, 0.01\% and 1\% packet loss probability. Horizontal bars denote the TAT in each case.}
 \label{fig:load_over_time}
\end{figure}

We find that these overheads are negligible. This to due to our use of x86 SSE/AVX instructions and a GPU offload. When we use float16, performance doubles, as expected.

\smartparagraph{Switch resources.}
We comment on \system's usage of switch resources in relation to network bandwidth and number of workers.
As discussed in Appendix \ref{app:impl}, our implementation only makes use of the switch ingress pipeline in maximizing the number of vector elements that is processed at line rate.
Aggregation bandwidth affects the required pool size. We verified that the memory requirement is less than 10\% of switch resources.
The number of workers does not influence the resource requirements to perform aggregation at line rate. That said, the number of switch ports and pipelines obviously pose a cap on how many directly-attached workers are supported. A single pipeline in our testbed supports 16-64 workers depending on network speed. We describe how to move beyond a single rack scale in the next section.

\section{Extensions}
\label{sec:disc}

\smartparagraph{Scaling beyond a rack.}
We described \system in the context of a rack. However, large scale ML
jobs could span beyond a single rack. \system's design can support
multiple racks by hierarchically composing several instances of our
switch logic, although we do not have a testbed large enough to test
(or require) such a design. Each worker is connected to a top-of-rack
switch, which aggregates updates from the workers in the rack. Rather
than broadcast the result packet to the workers, it instead sends it
to a tier-1 aggregation switch, which aggregates updates from multiple
racks. This can continue with as many levels as are needed to support
the desired network topology. Ultimately, a root switch completes the
aggregation of partial aggregates and multicasts a result packet
downstream. At each level, the switches further multicast the packet,
ultimately reaching the workers.

The hierarchical approach also allows us to support switches with
multiple processing pipelines. Suppose a switch has pipelines that can
aggregate up to $p$ ports (for the switches we use, $p=16$). In this
setting, each switch aggregates tensors from $d$ downstream ports and
forwards partial aggregates via $u = \lceil \frac{d}{p} \rceil$
upstream ports. In other words, the switch
operates as $u$ virtual switches, one for each pipeline in the
switch.

This hierarchical composition is bandwidth-optimal, as
it allows $n$ workers to fully utilize their bandwidth while
supporting all-to-all communication with a bandwidth cost proportional to
$u$ instead of $n$. That is, every switch aggregates data in a $p:1$
ratio. As a result, the system naturally supports oversubscription of
up to this ratio at the aggregation or core layers. This allows it to
support large clusters with relatively shallow hierarchies; using the
current generation of 64-port, 4-pipeline 100G switches, a two-layer
hierarchy can support up to 240 workers and a three-layer one up to
3600.

Importantly (and by design), the packet loss recovery algorithm
described in \S\ref{sec:packet-loss} works in the multi-rack
scenario. Thanks to the use of bitmaps and shadow copies, a
retransmission originated from a worker will be recognized as a
retransmission on any switches that have already processed that
packet. This triggers the retransmission of the updated value toward
the upper layer switch, ensuring that the switch affected by the
packet loss is always ultimately reached.

\smartparagraph{Congestion control.} We have not implemented an
explicit congestion control algorithm; the self-clocking streaming
protocol is a flow control mechanism to control access to the switch's
aggregator slots. It also serves as a rudimentary congestion control
mechanism, in that if one worker's link is congested and it cannot
process aggregation results at full speed, the self-clocking mechanism
will reduce the sending rate of all workers. This is sufficient for dedicated networks (which is common for ML clusters in practice).
For more general use, a congestion control scheme may
be needed; concurrent work has been developing such
protocols~\cite{panama}.

\smartparagraph{Deployment model.}
Thus far, we presented \system as an in-network computing approach, focusing on the mechanisms to enable efficient aggregation of model updates at line rate on programmable switching chips with very limited memory.
While that might be a viable deployment model in some scenarios, we highlight that our design may have more ample applicability. In fact, one could use a similar design to create a dedicated ``parameter aggregator,'' i.e., a server unit that combines a programmable switching chip with a typical server board, CPU and OS. Essentially a standard server with an advanced network attachment, or in the limit, an array of programmable Smart NICs, each hosting a shard of aggregator slots. The switch component of \system would run on said network attachment.
Then, racks could be equipped with such a parameter aggregator, attached for example to the legacy ToR using several 100 Gbps or 400 Gbps ports, or via a dedicated secondary network within the rack directly linking worker servers with it. We expect this would provide similar performance improvements while giving more options for deployment configurations; concurrent work has been exploring a similar approach atop an FPGA board~\cite{netreduce}.

\smartparagraph{Multi-job (tenancy).}
In multi-job or multi-tenant scenarios, the question arises as to how to support concurrent reductions with \system.
The solution is conceptually simple. Every job requires a separate pool of aggregators to ensure correctness. As discussed, the resources used for one reduction are much less than 10\% of switch capabilities. Moreover, modern switch chips comprise multiple independent pipelines, each with its own resources. Thus, an admission mechanism would be needed to control the assignment of jobs to pools.

\smartparagraph{Encrypted traffic.}
Given the cluster setting and workloads we consider, we do not consider it necessary to accommodate for encrypted traffic. Appendix \ref{app:enc} expands on this issue.

\smartparagraph{Asynchronous SGD.}
As mentioned, we only target synchronous SGD, since it is commonly used in the cluster setting to enable reproducibility. 

\section{Related work}
\label{sec:related}

\smartparagraph{In-network computation trends.}
The trend towards programmable data planes
has sparked a surge of proposals~\cite{netpaxos, paxos-switch-y, netchain, incbricks, netcache, eris}
to offload, 
when appropriate~\cite{when-in-net}, 
application-specific primitives into network devices.

\smartparagraph{In-network aggregation.}
We are not the first to propose aggregating data in the network.
Targeting partition-aggregate and big data (MapReduce) applications, NetAgg~\cite{netagg} and CamDoop~\cite{camdoop} demonstrated significant performance advantages, by performing application-specific data aggregation at switch-attached high-performance middleboxes or at servers in a direct-connect network topology, respectively. Parameter Hub~\cite{phub2018} does the same with a rack-scale parameter server.
Historically, some specialized supercomputer networks ~\cite{bluegenel,bluegenep} offloaded MPI collective operators (e.g., all-reduce) to the network.
\system differs from all of these approaches in that it performs in-network data reduction using a streaming aggregation protocol.

The closest work to ours is DAIET~\cite{daiet}. The authors also proposed in-network aggregation for minimizing communication overhead of exchanging ML model updates. However, their short paper does not describe a complete design, does not address the major challenges (\S\ref{sec:chall}) of supporting ML applications, and provides only a simple proof-of-concept prototype for MapReduce applications running on a P4 emulator. 
It is unclear whether it could work on a real switch.

Mellanox's Scalable Hierarchical Aggregation Protocol (SHARP) is a
proprietary in-network aggregation scheme available in certain
InfiniBand switches~\cite{sharp}. SHARP uses dedicated
on-chip FPUs for collective offloading. The most recent version,
SHARPv2~\cite{sharpv2} uses streaming aggregation analogous
to ours. A key difference is that SHARP builds on InfiniBand where it
can leverage link-layer flow control and lossless guarantees,
whereas SwitchML runs on standard Ethernet\footnote{Although SwitchML
  uses RDMA, it uses only unreliable connections, and so does
  \emph{not} require any of the ``lossless Ethernet'' features of
  RoCE.} with an unmodified network architecture, necessitating a new
packet recovery protocol. More fundamentally, SwitchML builds on
programmable network hardware rather than SHARP's fixed-function FPUs,
which offers two benefits. First, operators can deploy a single
switch model either for SwitchML or traditional networking without waste:
the ALUs used for aggregation can be repurposed for other tasks.
Second, it allows the system design to evolve to support new ML
training approaches. For example, we are
currently experimenting with new floating-point representations and
protocols for sparse vector aggregations. With a fixed-function
approach, these would require new hardware, just as moving from single
HPC reductions (SHARPv1) to streaming ML reductions (SHARPv2)
required a new ASIC generation.

Concurrently, Li et al.~\cite{iswitch} explored the idea of in-switch acceleration for Reinforcement Learning (RL). Their design (iSwitch) differs from ours in two fundamental ways. First, while their FPGA-based implementation supports more complex computation (native floating point), it operates at much lower bandwidth (4x10~Gbps). 
Second, it stores an entire gradient vector during aggregation; for RL workloads with small models, this works, but it does not scale for large DNN models. Our work targets both large models and high throughput -- a challenging combination given the limited on-chip memory in high-speed networking ASICs.
\system's software/hardware co-design approach, using a self-clocking streaming protocol, provides $40\times$ higher throughput than iSwitch, while supporting arbitrarily large models.

\smartparagraph{Accelerating DNN training.}
A large body of work has proposed improvements to hardware and
software systems, as well as algorithmic advances for faster DNN training. We only discuss a few relevant prior approaches.
Improving training performance via data or model parallelism has been explored by numerous deep learning systems \cite{distbelief, adam, tensorflow, bosen, mxnet, scaling-with-ps, phub2018}.
While data parallelism is most common, it can be advantageous to combine
the two approaches. Recent work even shows how to automatically find a fast parallelization strategy for a specific parallel machine \cite{flexflow}.
Underpinning any distributed training strategy, lies parameter synchronization.
Gibiansky was among the first to research \cite{gibiansky-gtc2017}
using fast collective algorithms in lieu of the traditional parameter server approach.
Many platforms have now adopted this approach~\cite{firecaffe,gibiansky-gtc2017,applied-ML-FB,horovod,cntk}.
We view \system as a further advancement on this line of work -- one that pushes the boundary by co-designing networking functions with ML applications.

\section{Conclusion}

\system speeds up DNN training by minimizing communication overheads at single-rack scale. \system uses in-network aggregation to efficiently synchronize model updates at each training iteration among distributed workers executing in parallel.
We evaluate \system with eight real-world DNN benchmarks on a GPU cluster with 10 Gbps and 100 Gbps networks; we show that \system achieves training throughput speedups up to 5.5$\times$ and is generally better than state-of-the-art collective communications libraries.

\label{lastpage}

\bibliographystyle{abbrv}
\bibliography{main}

\appendix

\begin{figure*}[t]
  \centering
  \includegraphics[width=0.9\textwidth]{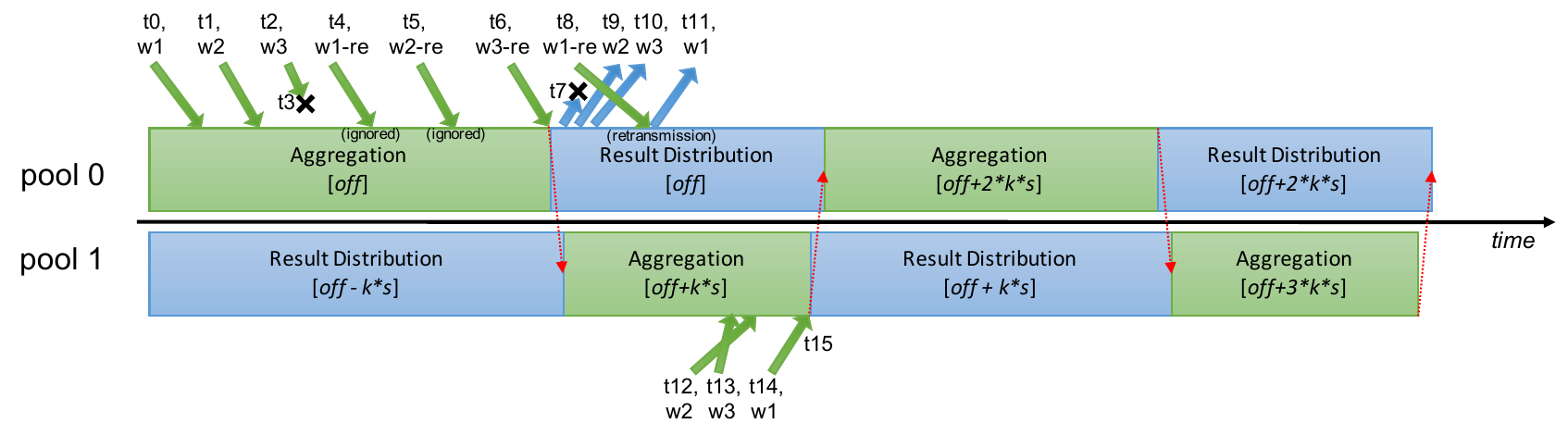}
  \caption{An example execution of a \system switch interacting with three workers. The figure illustrates how a slot with index $x$ is used during the different phases (shown in different colors) that alternate between the two pools.}
  \label{fig:example}
\end{figure*}

\section{Example execution}
\label{app:example}

We illustrate through an example how Algorithms \ref{algo:switch_prog_loss} and \ref{algo:worker_prog_loss} behave in a system with three workers: w1, w2, and w3. We focus on the events that occur for a particular slot ($x$) starting from a particular offset ($\textit{off}$).

\begin{itemize}
\item {\bf t0:} Worker w1 sends its model update for slot $x$ with offset = $\textit{off}$.
\item {\bf t1:} Worker w2 sends its model update for slot $x$ with offset = $\textit{off}$.
\item {\bf t2:} Worker w3 sends its model update for slot $x$ with offset = $\textit{off}$. This update packet is lost on the upstream path at time {\bf t3}, and hence the switch does not receive it.
\item {\bf t4:} w1's timeout kicks in for the model update sent at {\bf t0}, leading to a retransmission of the same model update for slot $x$ with offset = $\textit{off}$. The switch receives the packet, but it ignores the update because it already received and aggregated an update from w1 for the given slot and offset.
\item {\bf t5:} w2's timeout kicks in for the model update sent at {\bf t0}, leading to a retransmission of the same model update for slot $x$ with offset = $\textit{off}$. The switch receives the packet, but it ignores the update because it already received and aggregated an update from w2 for the given slot and offset.
\item {\bf t6:} w3's timeout kicks in for the model update sent at {\bf t0}, leading to a retransmission of the same model update for slot $x$ with offset = $\textit{off}$. The switch receives the packet and aggregates the update properly. Since this update is the last one for slot $x$ and offset $\textit{off}$, the switch completes aggregation for the slot and offset, turns the slot into a shadow copy, and produces three response packets (shown as blue arrows). 
\item {\bf t7:} The first response packet for w1 is lost on the downstream path, and w1 does not receive it. 
\item {\bf t8:} Not having received the result packet for the update packets sent out earlier (at {\bf t0} and {\bf t4}), w1 retransmits its model update the second time. This retransmission reaches the switch correctly, and the switch responds by sending a unicast response packet for w1.
\item {\bf t9} and {\bf t10:} w2 and w3 respectively receives the response packet. Hence, w2 and w3 respectively decides to reuse slot $x$ for the next offset ($\textit{off} + k \cdot s$) and sends their new updates at {\bf t12} and {\bf t13}.
\item {\bf t11:} The unicast response packet triggered by the second model-update retransmission (sent at {\bf t8}) arrives at w1. 
\item {\bf t14:} Now that w1 has received its response, it decides to reuse slot $x$ for the next offset ($\textit{off} + k \cdot s$) and sends its new updates. This update arrives at the switch at {\bf t15}, upon which the switch realizes that the slot for offset ($\textit{off} + k \cdot s$) is complete. This confirms that the result in the shadow-copy slot (the slot in pool 0) is safely received by every worker. Thus, the switch flips the roles of the slots again.
\end{itemize}

\section{Implementation details}
\label{app:impl}

\smartparagraph{Switch component.}
The main challenge we faced 
was to find a design that best utilizes the available resources (SRAM, TCAMs, hashing machinery, etc.) to perform as much computation per packet as possible. Data plane programs are typically constrained by either available execution resources or available storage; for \system, execution resources are the tighter constraint. For example, a data plane program is constrained by the number of stages per pipeline~\cite{domino}, which limits the dependencies within the code. In fact, every action whose execution is contingent on the result of a previous operation has to be performed on a subsequent stage. A program with too many dependencies cannot find a suitable allocation on the hardware pipeline and will be rejected by the compiler. Moreover, the number of memory accesses per-stage is inherently limited by the maximum per-packet latency; a switch may be able to parse more data from a packet than it is able to store into the switch memory during that packet's time in the switch. 

We make a number of design trade-offs to fit within the switch constraints. First, our P4 program makes the most use of the limited memory operations by performing the widest register accesses possible (64 bits). We then use the upper and lower part of each register for alternate pools. These parts can execute different operations simultaneously; for example, when used for the received work bitmap, we can set a bit for one pool and clear a bit for the alternate pool in one operation. Second, we minimize dependencies (e.g., branches) in our Algorithm~\ref{algo:switch_prog_loss} in order to process 64 elements per packet within a single ingress pipeline. We confine all processing to the ingress pipeline; when the aggregation is complete, the traffic manager duplicates the packet containing the aggregated result and performs a multicast. In a first version of our program, we used both ingress and egress pipelines for the aggregation, but that required packet recirculation to duplicate the packets. This caused additional dependencies that required more stages, preventing the processing of more than 64 elements per packets. Moreover, this design experienced unaccounted packet losses between the two pipelines and during recirculation, which led us to search for a better, single pipeline, program.

\smartparagraph{Worker component.}
Our goal for implementing the worker component is to achieve high I/O performance for aggregating model updates. At the same time, we want to support existing ML frameworks without modifications.

In existing ML frameworks, a DNN model update $U$ comprises of a set of tensors ${T}$, each carrying a subset of the gradients. This is because the model consists of many layers; most existing frameworks emit a gradient tensor per layer and reduce each layer's tensors independently.
Back-propagation produces the gradients starting from the output layer and moving towards the input layer. Thus, communication can start on the output layer's gradients while the other gradients are still being computed, partially overlapping communication with computation.
This implies that for each iteration, there are as many aggregation tasks as the number of tensors (e.g., 152 for ResNet50).

Our implementation exposes the same synchronous all-reduce interface as Gloo. However, rather than treating each tensor as an independent reduction and resetting switch state for each one, our implementation is efficient in that it treats the set of tensors virtually as a single, continuous stream of data across iterations. Upon invocation, our API passes the input tensor to a virtual stream buffer manager which streams the tensor to the switch, breaking it into the small chunks the switch expects. Multiple threads may call \system's all-reduce, with the requirement that each worker machine's tensor reduction calls must occur in the same order; the stream buffer manager then performs the reductions and steers results to the correct requesting thread.

One CPU core is sufficient to do reduction at line rate on a 10 Gbps network. However, to be able to scale beyond 10 Gbps, we use multiple CPU cores at each worker and use the Flow Director technology (implemented in hardware on modern NICs) to uniformly distribute incoming traffic across the NIC RX queues, one for each core.
Every CPU core runs an I/O loop that processes every batch of packets in a run-to-completion fashion and uses a disjoint set of aggregation slots. Packets are batched in groups of 32 to reduce per-packet transmission overhead.
We use x86 SSE/AVX instructions to scale the model updates and convert between types.
We are careful to ensure all processing is NUMA aware.

\paragraph{RDMA implementation details.}
\label{sec:rdmadetails}
We found that the cost of processing individual SwitchML packets, even using DPDK with 256-element packets and multiple cores, was too high to achieve line rate. Other aggregation libraries use RDMA to offload packet processing to the NIC. In RDMA-based systems NICs implement packetization, flow control, congestion control, and reliable delivery. In normal usage, clients use RDMA to send and receive messages of up to a gigabyte; the NIC turns them into packets and ensures they are delivered reliably. Furthermore, clients may register memory regions with the NIC, allowing other clients to remotely read and write them without CPU involvement. This reduces or eliminates work done on the clients' CPUs to complete the transfer.

Turning a Tofino switch into a fully-featured RDMA endpoint is not the solution. Implementing timeouts and retransmission in a way that is compatible with existing poorly-documented existing RDMA NICs would be complex. Furthermore, such an implementation would not be an good fit for SwitchML: the RDMA protocols are largely designed for point-to-point communication, whereas SwitchML's protocol is designed for collective communication.

Fortunately, RDMA NICs implement multiple protocols with different properties. The standard Reliable Connected (RC) mode ensures reliable delivery and supports CPU-bypassing remote reads and writes (as well as sends and receives) of up to 1 GB. The UDP-like Unreliable Datagram (UD) mode supports just sends and receives of up the network MTU. Finally, the Unreliable Connected (UC) mode fits somewhere in between. It supports packetization, allowing for sends, receives, and writes of up to 1 GB. It also generates and checks sequence numbers, allowing it to detect packet drops, but it does not retransmit: instead, if a gap in sequence numbers is detected, incoming packets are silently dropped until the first packet of a new message arrives. Then, the sequence number counter is reset to the sequence number of that packet, and normal reception continues.

We use RDMA UC to implement a RDMA-capable variant of SwitchML using a subset of the RoCE v2 protocol~\cite{rocev2}. Its operation is very similar to what is described in Section~\ref{sec:design}, with three main differences. 

First, where base SwitchML sends and receives slot-sized packets, SwitchML RDMA sends multi-slot messages. Each packet of a message is treated largely as it is in the base protocol by the switch, but the pool index for each packet is computed as an offset from the base index provided with the first packet of the message. Timeouts are tracked just as they are in the base protocol, but when a packet drop is detected, the client retransmits the entire message rather than just the dropped packet. This makes retransmissions more expensive, but it also drastically lowers the cost incurred sending packets in the common case of no packet drops; since packet drops are rare within a datacenter, the benefit is large.

Second, SwitchML consumes and generates sequence numbers on the switch. In order to allow messages with multiple packets to aggregate concurrently, each in-flight message is allocated its own queue pair, with its own sequence number register. This allows clients to to be notified when a write message from the switch has arrived with no drops; it also allows the switch to ignore packets in messages received out of sequence.
However, the same per-slot bitmap used in the base protocol is still used to ensure that duplicate packets from a retransmission of a partially-received messages are not re-applied. Packets are transmitted as individual slots addressed by a message complete. This means that the packets from multiple messages may interleave on the wire, but since each is on a separate queue pair with its own sequence number space, the NIC will reassemble them successfully.

Third, SwitchML RDMA uses RDMA Write Immediate messages for all communication. This allows clients to send data directly from GPU memory, and the switch to write directly into GPU memory (if the host is GPU Direct-capable). Byte order conversion and scaling are done on the GPU; the CPU is responsible only for issuing writes when data from the GPU is ready, detecting completions and timeouts, and issuing retransmissions when necessary. Necessary metadata for the SwitchML protocol is encoded in fields of the RDMA header; the RDMA RKey and Address fields are used to encode the destination slot and the address to write the response to. The Immediate field is used to carry up to four scaling factors. At job setup time, the clients communicate with the switch and give it their queue pair numbers, initial sequence numbers, and an RKey for its switch-writable memory region. The switch uses these to form RDMA Write Immediate messages with appropriate sequence numbers, destination addresses, and immediate values, of the same size as the messages sent from the clients to the switch.

Finally, it is important to note that SwitchML RDMA does not require lossless Ethernet to be configured, as is common in RoCE deployments. Enabling lossless Ethernet would reduce the probability of packet drops, but would add complexity to the network deployment. SwitchML's reliability protocol makes this unnecessary.

\smartparagraph{DNN workloads.}
Table~\ref{tab:dnns-config} details the models, datasets and ML toolkits used in the experiments.

\begin{figure}[tp!]
 \centering
 \includegraphics[width=0.45\textwidth]{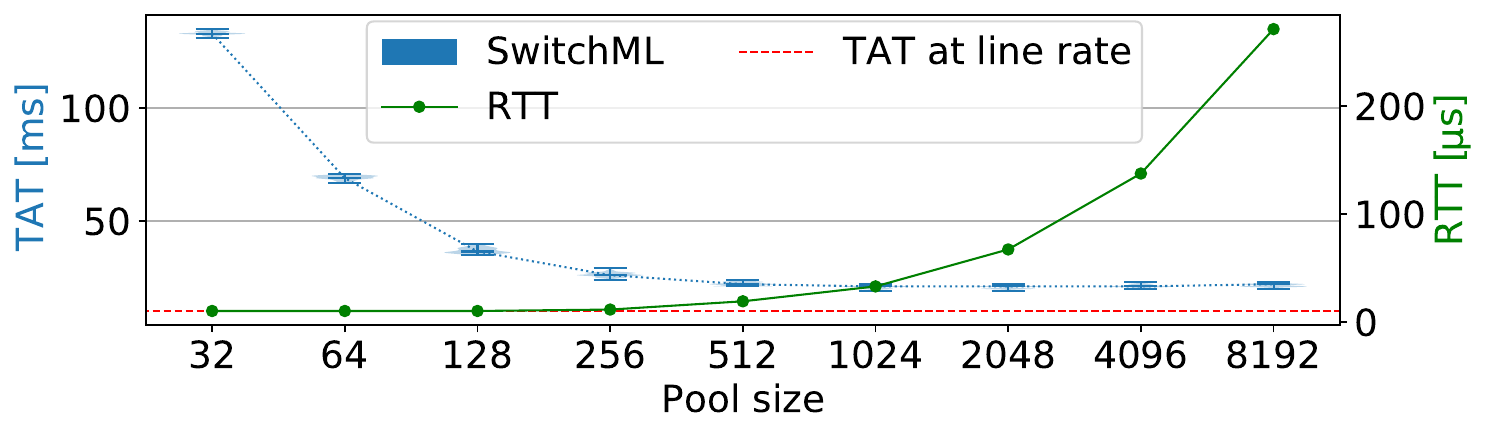}
 \caption{Effect of pool size on overall tensor aggregation time (TAT) and per-packet RTT (right y-axis) at 100 Gbps.}
 \label{fig:poolsize100G}
\end{figure}

\section{Tuning the pool size}
\label{sec:tuning}

As mentioned, the pool size $s$ affects performance and reliability. We now analyze how to tune this parameter.

Two factors affect $s$. First, because $s$ defines the number of in-flight packets in the system that originate from a worker, to avoid wasting each worker's network bandwidth, $s$ should be no less than the bandwidth-delay product (BDP) of each worker. Note that the delay here refers to the end-to-end delay, including the end-host processing time, which can be easily measured in a given deployment. Let $b$ be the packet size, which is constant in our setting. To sustain line rate transmission, the stream of response packets must arrive at line rate, and this is possible when $s \cdot b$ matches the BDP. A significantly higher value of $s$, when used as the initial window size, will unnecessarily increase queuing time within the workers.

Second, a correctness requirement for our communication scheme is that no two in-flight packets from the same worker use the same slot (as no worker node can ever lag behind by more than one phase). 
To sustain line rate and preserve correctness, the lower bound on $s$ is such that $s \cdot b$ matches the BDP. Therefore, the optimal $s$ is for $\lceil BDP / b \rceil$. 

In practice, we select $s$ as the next power of two of the above formula because the DPDK library -- which we use to implement \system~-- performs batched send and receive operations to amortize system overheads.
Based on our measurements (Figure \ref{fig:poolsize100G}), we use 128 and 512 as the pool size for 10 and 100 Gbps, respectively. This occupies 256 KB and 1 MB of register space in the switch, respectively. We note that the switch can support one order of magnitude more slots, and \system uses much less than $10\%$ of that available. 

\begin{figure*}[t]
  \centering
  \includegraphics[width=0.9\textwidth]{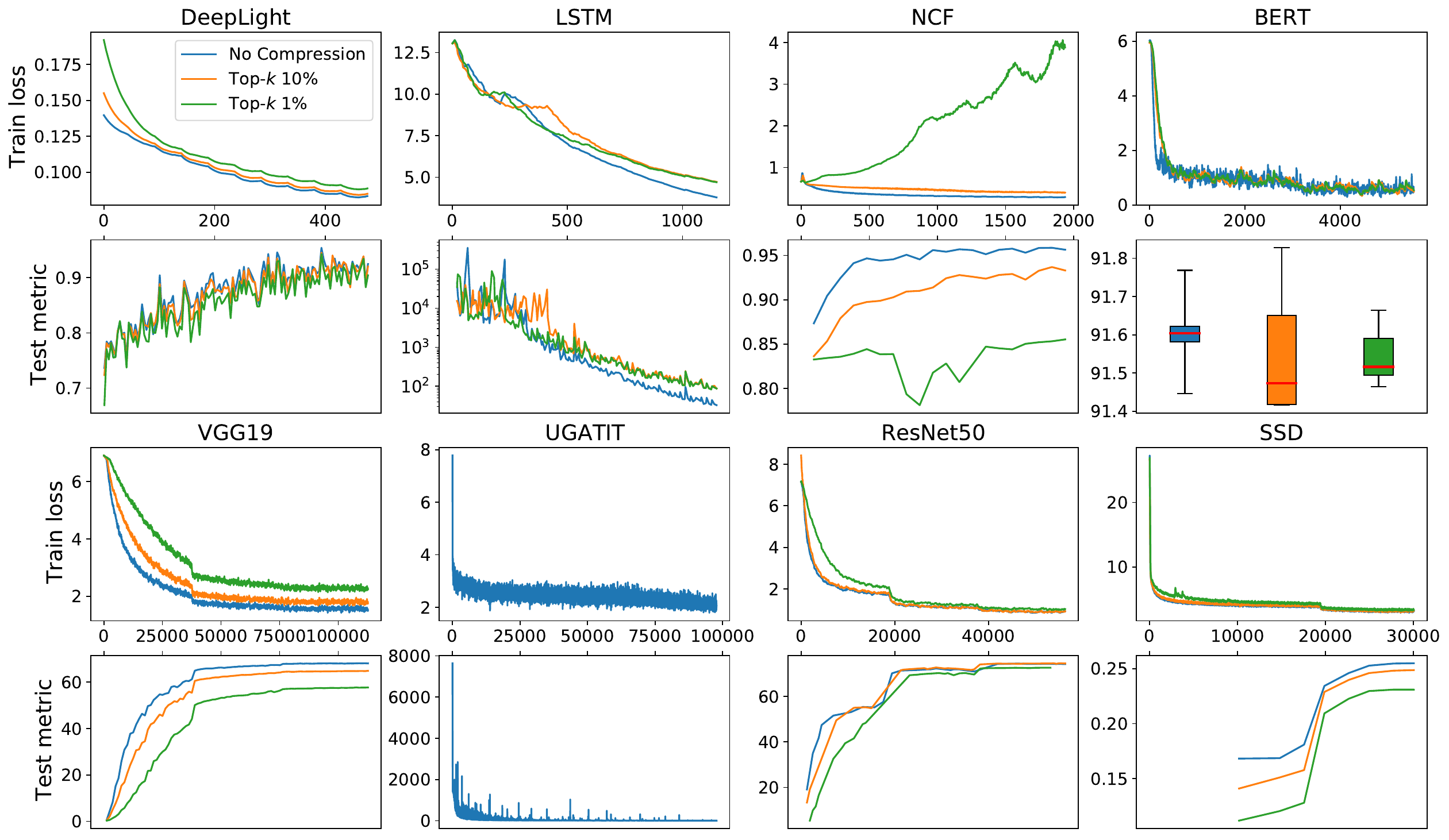}
  \caption{Convergence behavior under different compression scheme. The x axis shows the iteration number. All methods execute a fixed number of steps and hyperparameters are kept the same. Refer to Table \ref{tab:convergence} for test metrics of each task. We plot generator loss as the metric for UGATIT because there is no objective metric available for GAN workloads. Note that for the perplexity metric of LSTM, lower is better.}
  \label{fig:convergence}
\end{figure*}

\begin{figure*}[t]
  \centering
  \includegraphics[width=0.9\textwidth]{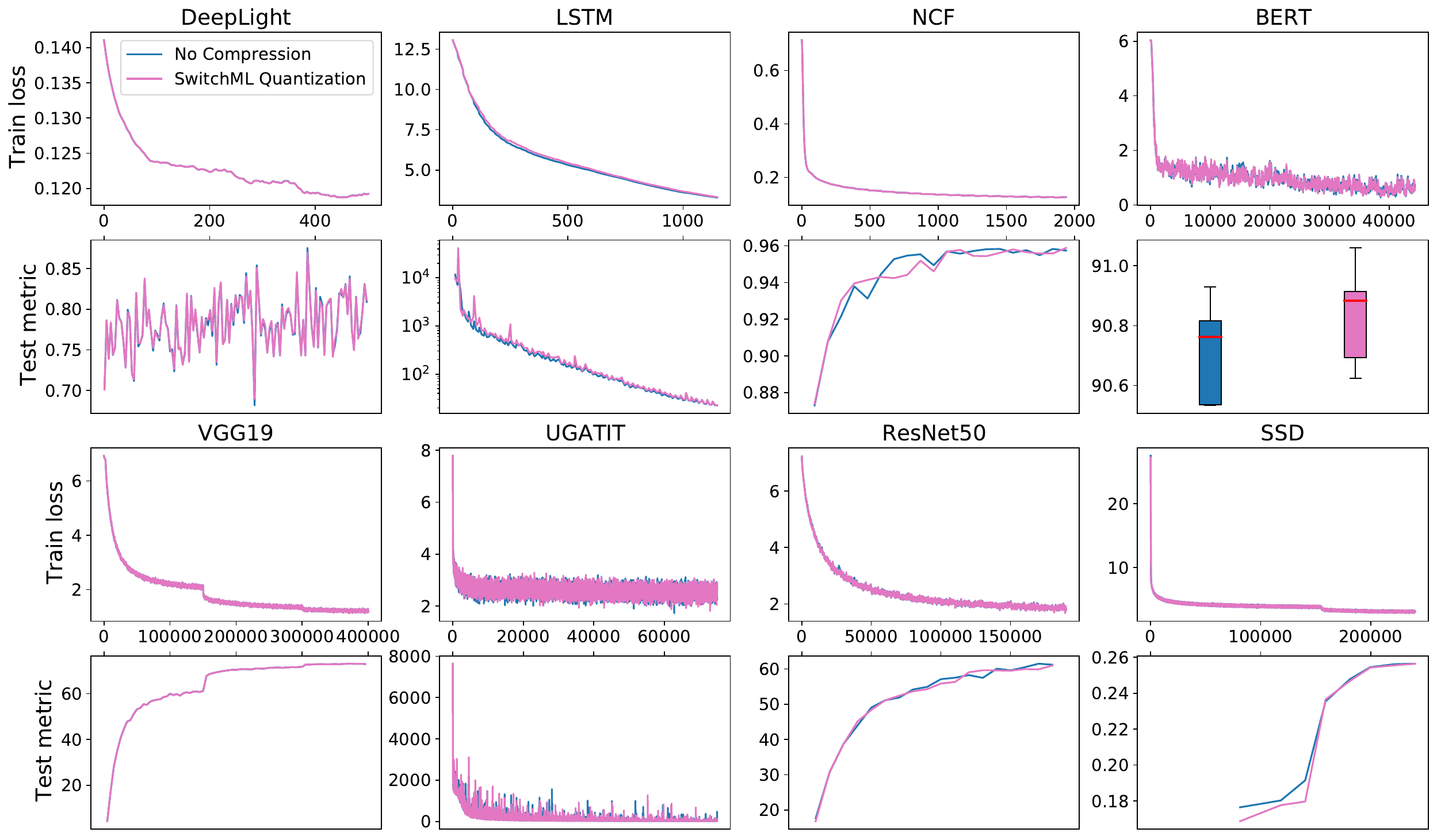}
  \caption{Convergence analysis of SwitchML quantization method on a single GPU. The x axis shows the iteration number.}
  \label{fig:convergence_switchml}
\end{figure*}

\begin{table}[t]
\resizebox{\columnwidth}{!}{%
\begin{tabular}{|l|l|l|r|r|}
\hline
\textbf{Model} & \textbf{Task} & \textbf{Dataset} & \textbf{Sys} \\ \hline
DeepLight~\cite{Deng2020DeepLightDL}      & Click-through Rate Prediction  & Criteo 1TB~\cite{criteo} & PyT  \\ \hline
LSTM~\cite{jozefowicz2016exploring}      & Language Modeling & GBW~\cite{chelba2013one} & PyT \\ \hline
NCF~\cite{ncf}            & Recommendation & ML-20m~\cite{harper2015movielens} & PyT  \\ \hline
BERT~\cite{devlin2018bert}           & Question Answering &  SQuAD~\cite{rajpurkar2018know} & TF \\ \hline
VGG19~\cite{vgg}          & Image Classification &  ImageNet-1K~\cite{imagenet} & PyT \\ \hline
UGATIT~\cite{Kim2020U-GAT-IT:}      & Image-to-Image Translation &  Selfie2Anime~\cite{Kim2020U-GAT-IT:} & TF \\ \hline
ResNet50~\cite{resnet}      & Image Classification &  ImageNet-1K~\cite{imagenet} & TF \\ \hline
SSD~\cite{liu2016ssd}      & Object Detection &  COCO 2017~\cite{DBLP:journals/corr/LinMBHPRDZ14} & PyT \\ \hline
\end{tabular}
}
\caption{DDL benchmarks. Models, task, dataset used for training and ML toolkit (PyT=PyTorch; TF=TensorFlow).}
\label{tab:dnns-config}
\end{table}

\section{Compression affects convergence}
\label{app:compression}

Table~\ref{tab:convergence} reports the model quality obtained without gradient compression and with Top-$k$ compression using $k=1\%,10\%$.
Model quality is assessed using per-model accuracy metrics.
  
Results are show in Figure~\ref{fig:convergence}. We observe that loss and accuracy do not necessary correlate well. For example, in the case of SSD all methods have similar loss trace, but obvious accuracy gap. For NCF, Top-$k$ at 1\% does not converge, but the accuracy can still go up.

\begin{table}[t]
\resizebox{\columnwidth}{!}{%
\begin{tabular}{|l|r|r|r|r|}
\hline
\textbf{Model}      & \textbf{Metric} &  \textbf{No compression} & \textbf{Top-10\%} & \textbf{Top-1\%}   \\ \hline
DeepLight  & AUC & 0.9539 & 0.9451 & 0.9427   \\ \hline
LSTM        & Perplexity   & 32.74 & 86.26 & 82.55  \\ \hline
NCF        & Hit rate   & 0.9586 & 0.9369 & \xxmark  -  \\ \hline
BERT        & F1 score & 91.60 & 91.47 & 91.52   \\ \hline
VGG19      & Top-1 accuracy & 68.12  & 64.85  & 57.70    \\ \hline
UGATIT & \xxmark - & \xxmark - & \xxmark -&  \xxmark - \\\hline
ResNet50      & Top-1 accuracy & 74.34  & 74.59  & 72.63 \\ \hline
SSD        & Accuracy   & 0.2549 & 0.2487 & 0.2309       \\
\hline
\end{tabular}
}
  \caption{Test metrics comparison. NCF at Top-1\% did not converge. BERT result is the median of 6 runs of fine-tuning from a pre-trained model. UGATIT fails to execute with the compressor implementation in~\cite{grace}. See Figure \ref{fig:convergence} for the convergence behavior during training.}
  \label{tab:convergence}
\end{table}

\section{Model quantization}
\label{app:quant}

To the best of our knowledge, no Ethernet switching chip offers floating-point operations in the dataplane for packet processing. Some InfiniBand switching chips have limited support for floating-point operations for scientific computing~\cite{sharp}. We also confirmed that the state-of-the-art programmable Ethernet switching chips do not support native floating-point operations either. These observations lead us to two main questions.

\begin{figure}[t]
 \centering
 \includegraphics[width=0.45\textwidth]{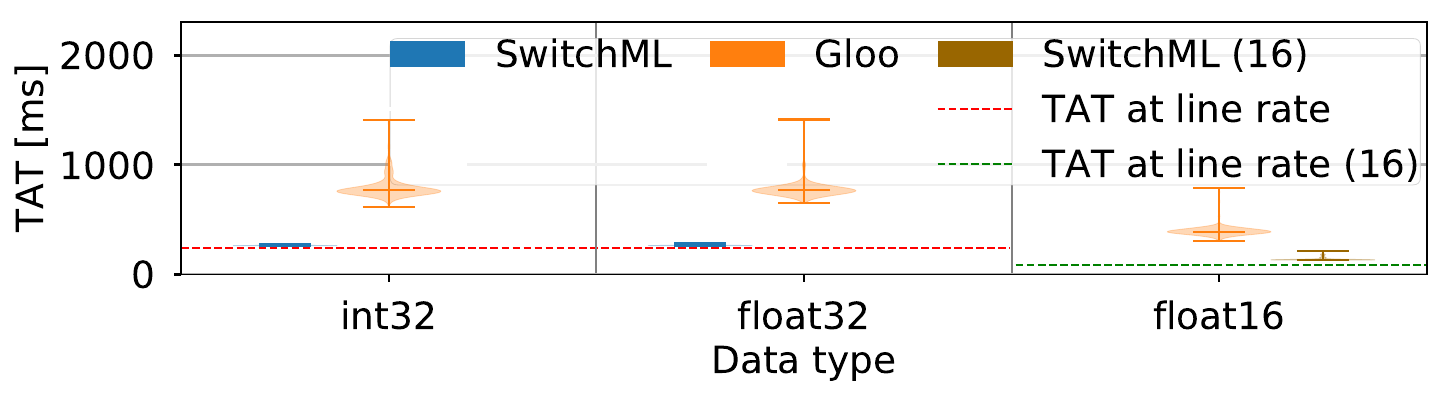}
 \caption{TAT comparison between aggregating native-integer tensors, scaling and converting from single-precision (float32) tensors, and scaling and converting from half-precision (float16) tensors.}
 \label{fig:data_type}
\end{figure}

\smartparagraph{Where should the type conversion occur?}
In theory either workers or the switch can perform type conversion.
In the former case, packets carry a vector of integer types while in the latter case the switch internally performs the type conversions.
It turns out to be possible to implement a restricted form of 16-bit floating point on a Tofino chip by using lookup tables and ALUs to do the conversion. This means there is no type conversion overhead at end hosts. However, this approach still requires to scale the gradient values due to the limited range of floating point conversion the switch can perform. Besides, unless half-precision training is used, the worker must still convert from 32-bit to 16-bit floating points.
At the same time, an efficient implementation that uses modern x86 vector instructions (SSE/AVX) to implement type conversion sees only a negligible overhead (see Figure \ref{fig:data_type}).

\smartparagraph{What are the implications in terms of accuracy?}
Recently, several  update compression (e.g., quantization, dithering or sparsification) strategies were proposed to be used with standard training methods, such as SGD, and bounds were obtained on how these strategies influence the number of iterations until a sufficient convergence criterion is satisfied (e.g., being sufficiently close to minimizing the empirical loss over the
data set). These include aggressive 1-bit compression of SGD for DNNs \cite{seide2014}, signSGD \cite{bernstein2018signsgd, pmlr-v80-bernstein18a}, QSGD \cite{alistarh2017qsgd}, which uses just the sign of the stochastic gradients to inform the update, Terngrad \cite{wen2017terngrad}, which uses ternary quantization, and the DIANA framework \cite{DIANA}, which allows for a wide array of compression strategies applied to gradient differences. All the approaches above use lossy randomized compression strategies that preserve unbiasedness of the stochastic gradients at the cost of increasing the variance of gradient estimators, which leads to worse iteration complexity bounds. Thus, there is a trade-off between savings in communication and the need to perform more training rounds. In contrast, our mechanism is not randomized, and for a suitable selection of a scaling parameter $f$, is essentially lossless or suffers negligible loss only.

We shall now briefly describe our quantization mechanism. Model updates are divided into packet-sized blocks. With a small abuse of notation, in the following all equations refer to per-block operations. Each worker multiplies its block model update $\Delta_i^t = \Delta(x^t,D_i^t)$ by a vector of scaling factors $f>0$, obtaining  $f \Delta_i^t$. The elements of $f$ are chosen per block, and are chosen such that that all $k$ entries of the scaled update can be rounded to a number representable as an integer without overflow. We then perform this rounding, obtaining vectors $Q_i^t = \rho(f \Delta_i^t) \in \mathbb{Z}^k$ for $i=1,2,\dots,n$, where $\rho$ is the rounding operator, which are sent to the switch and aggregated, resulting in
\[A^t = \sum_{i=1}^n Q_i^t.\]
Again, we need to make sure $f$ is not too large so that $A^t$ can be represented as an integer without overflow. The aggregated update $A^t$ is then scaled back on the workers, obtaining $A^t/f$, and the model gets updated as follows:
\[x^{t+1} = x^t + \frac{1}{f}A^t.\]

Let us illustrate this on a simple example with $n=2$ and $d=1$. Say $\Delta^t_1 = 1.56$ and $\Delta^t_2 = 4.23$. We set $f=100$ and get \[Q^t_1 = \rho(f \Delta^t_1) = \rho(156)= 156\] and \[Q^t_2 = \rho(f \Delta^t_2) = \rho(423) = 423.\] The switch will add these two integers, which results in $A^t = 579$. The model then gets updated as:
\[x^{t+1}  =x^t + \frac{1}{100} 579 = x^t + 5.79.\]
Notice that while the aggregation was done using integers only (which is necessitated by the limitations of the switch), the resulting model update is identical to the one that would be applied without any conversion in place. Let us consider the same example, but with $f=10$ instead. This leads to 
\[Q^t_1 = \rho(f \Delta^t_1) = \rho(15.6)= 16\] and \[Q^t_2 = \rho(f \Delta^t_2) = \rho(42.3) = 42.\] The switch will add these two integers, which results in $A^t = 58$. The model then gets updated as:
\[x^{t+1}  =x^t + \frac{1}{10} 58 = x^t + 5.8.\]
Note that this second approach leads to a small error. Indeed, while the true update is $5.79$, we have applied the update $5.8$ instead, incurring the error  $0.01$.

Our strategy is to apply the above trick, but  take special care about how we choose the scaling factor $f$ so that the trick works throughout the entire iterative process with as little information loss as possible.

\smartparagraph{A formal model.} Let us now formalize the above process. We first assume that we have a scalar $f>0$ for which the following holds:

{\em Assumption 1. $|\rho(f \Delta^t_i)| \leq 2^{31}$ for all $i=1,2,\dots,n$ and all iterations $t$.}

{\em Assumption 2. $|\sum_{i=1}^n \rho(f \Delta^t_i)| \leq 2^{31}$ for all  iterations $t$.}

The above assumptions postulate that all  numbers which we obtain by scaling and rounding on the nodes (Assumption 1), and by aggregation on the switch (Assumption 2), can be represented as integers without overflow.

We will now establish a formal statement which characterizes the error incurred by our aggregation procedure.

{\em Theorem 1 (Bounded aggregation error).  The difference between the exact aggregation value $\sum_{i=1}^n \Delta^t_i$ (obtained in case of perfect arithmetic without any scaling and rounding, and with a switch that can aggregate floats) and  the value $\frac{1}{f}A^t= \frac{1}{f}\sum_{i=1}^n \rho(f \Delta_i^t)$ obtained by our procedure is bounded by $\frac{n}{f}$.}
\begin{proof}
To prove the above result, notice that
\begin{eqnarray*}
\frac{1}{f} \sum_{i=1}^n \rho(f \Delta^t_i) &\leq & \frac{1}{f} \sum_{i=1}^n \lceil f \Delta^t_i \rceil\\
&\leq & \frac{1}{f} \sum_{i=1}^n ( f \Delta^t_i +1) \\
&=& \left( \sum_{i=1}^n  \Delta^t_i \right) + \frac{n}{f}.
\end{eqnarray*}
Using the same argument,  we get a similar lower bound
\begin{eqnarray*}
\frac{1}{f} \sum_{i=1}^n \rho(f \Delta^t_i) &\geq & \frac{1}{f} \sum_{i=1}^n \lfloor f \Delta^t_i \rfloor\\
&\geq & \frac{1}{f} \sum_{i=1}^n ( f \Delta^t_i -1) \\
&=& \left( \sum_{i=1}^n  \Delta^t_i \right) - \frac{n}{f}.
\end{eqnarray*}
\end{proof}

Note that the error bound postulated in Theorem 1 improves as $f$ increases, and $n$ decreases. In practice, the number of nodes is constant $n=O(1)$. Hence, it makes sense to choose $f$ as large as possible while making sure Assumptions 1 and 2 are satisfied. Let us give one example for when these assumptions are satisfied. In many practical situations it is known that the model parameters remain bounded:\footnote{If desirable, this can  be enforced explicitly by the inclusion of a suitable hard regularizer, and by using projected SGD instead of plain SGD.} 

{\em Assumption 3. There exists $B>0$ such that $|\Delta_i^t|\leq B$ for all $i$ and $t$.}

As we shall show next,  if Assumption 3 is satisfied, then so is Assumption 1 and 2.

{\em Theorem 2 (No overflow).  Let Assumption 3 be satisfied. Then Assumptions 1 and 2 are satisfied (i.e., there is no overflow) as long as  $0<f\leq \frac{2^{31}-n}{nB}$. }

\begin{proof} We have $\rho(f \Delta_i^t) \leq f \Delta_i^t + 1 \leq f |\Delta_i^t| +1 \leq f B +1$. Likewise, $\rho(f \Delta_i^t) \geq f \Delta_i^t - 1 \geq - f |\Delta_i^t| -1 = -( f B +1)$. So, $|\rho(f \Delta_i^t)| \leq fB +1$. Hence, as soon as $0<f\leq \frac{2^{31}-1}{B}$, Assumption 1 is satisfied. This inequality is less restrictive as the one we assume. Similarly, $|\sum_i \rho(f \delta_i^t)| \leq \sum_i |\rho(f\Delta^t_i)| \leq \sum_i (fB +1) = n(fB+1)$. So, Assumption 2 is satisfied as long as $n(fB+1) \leq 2^{31}$, i.e., as long as $0<f \leq \frac{2^{31}-n}{nB}$.
\end{proof}

 We now put all of the above together. By combining Theorem 1 (bounded aggregation error) and Theorem 2 (no overflow), and if we choose $f = \frac{2^{31}-n}{nB}$, then the difference between the exact update $\sum_i \Delta^t_i$ and our update $\frac{1}{f}\sum_i \rho(f \Delta_i^t)$ is bounded by $\frac{n^2B}{2^{31}-n}$. In typical applications, $n^2 B \ll 2^{31}$, which means that the error we introduce is negligible.

\vspace{3em}
\section{Encrypted traffic}
\label{app:enc}

A recent trend, especially at cloud providers, is to encrypt all datacenter traffic. In fact, data encryption is generally performed at the NIC level itself.
While addressing this setting is out of scope, we wish to comment on this aspect.
We believe that given our substantial performance improvements, one might simply forego encryption for ML training traffic.

We envision a few alternatives for when that is not possible.
One could imagine using HW accelerators to enable in-line decryption/re-encryption at switches.
However, that is likely costly.
Thus, one may wonder if computing over encrypted data at switches is possible.
While arbitrary computations over encrypted data are beyond current switches' capabilities, we note that the operation performed at switches to aggregate updates is simple integer summation.
The appealing property of several partially homomorphic cryptosystems (e.g., Paillier) is that the relation $E(x) \cdot E(y) = E(x+y)$ holds for any two values $x,y$ ($E$ denotes encryption).
For instance, recent work by Cheon et al.~\cite{Cheon.ASIACRYPT2017} developed a homomorphic encryption scheme for approximate arithmetic.
By customizing the end-host encryption process, the worker could encrypt all the vector elements using such a cryptosystem, knowing that the aggregated model update can be obtained by decrypting the data aggregated at the switches.
We leave it to future work to validate this concept.

\end{document}